\documentclass[%
  aps,
  prb,
  twocolumn,
  superscriptaddress,
  longbibliography,
  amsmath,
  amssymb,
  floatfix
]{revtex4-2}

\usepackage{graphicx}
\usepackage[colorlinks=true,linkcolor=blue,citecolor=blue,urlcolor=blue]{hyperref}
\usepackage{graphicx}
\usepackage[utf8]{inputenc}

\usepackage{amsmath,latexsym}
\usepackage{xcolor}
\definecolor{Gray}{gray}{0.0}
\usepackage[%
  colorlinks=true,
  urlcolor=blue,
  linkcolor=blue,
  citecolor=blue
]{hyperref}
\usepackage{orcidlink}
\usepackage{xcolor}
\definecolor{mygreen}{RGB}{0,128,0}
\usepackage{amsmath}

\usepackage{libertine} 
\usepackage{siunitx}
\makeatletter
\DeclareSIUnit \electron {e}
\DeclareSIUnit \centi {c}
\newcommand*{\addFileDependency}[1]{
  \typeout{(#1)}
  \@addtofilelist{#1}
  \IfFileExists{#1}{}{\typeout{No file #1.}}
}

\usepackage[UKenglish]{babel}

\begin{document}

\title{Studying all-optical magnetization switching of GdFe by double-pulse laser excitation }

\author{\surname{Rahil} Hosseinifar\orcidlink{0000-0002-9124-008X}}%
\affiliation{Institut f\"ur Experimentalphysik, Freie Universit\"at Berlin, Arnimallee 14, 14195 Berlin, Germany}
\author{\surname{Felix} Steinbach\orcidlink{0000-0001-9287-1062}}%
\affiliation{Max-Born-Institut f\"ur Nichtlineare Optik und Kurzzeitspektroskopie, Max-Born-Stra{\ss}e 2A, 12489 Berlin, Germany}
\author{\surname{Ivar} Kumberg\orcidlink{0000-0002-3914-0604}}%
\affiliation{Institut f\"ur Experimentalphysik, Freie Universit\"at Berlin, Arnimallee 14, 14195 Berlin, Germany}
\author{\surname{José}  Miguel Lendínez\orcidlink{}}%
\affiliation{Instituto de Ciencia de Materiales de Madrid, CSIC, Cantoblanco, 28049 Madrid, Spain}
\author{\surname{Sangeeta} Thakur\orcidlink{0000-0003-4879-5650}}%
\author{\surname{Sebastien E.} Hadjadj\orcidlink{0000-0002-6045-574X}}%
\author{\surname{Jendrik } Gördes\orcidlink{}}%
\author{\surname{Chowdhury}  S. Awsaf\orcidlink{0009-0007-4709-6168}}%
\affiliation{Institut f\"ur Experimentalphysik, Freie Universit\"at Berlin, Arnimallee 14, 14195 Berlin, Germany}
\author{\surname{Mario} Fix\orcidlink{0000-0002-6677-2674}}%
\affiliation{Institute of Physics, University of Augsburg, Universit\"atsstra{\ss}e 1, 86135 Augsburg, Germany}
\author{\surname{Manfred} Albrecht\orcidlink{0000-0002-0795-8487}}%
\affiliation{Institute of Physics, University of Augsburg, Universit\"atsstra{\ss}e 1, 86135 Augsburg, Germany}
\author{\surname{Florian} Kronast\orcidlink{0000-0001-6048-480X}}%
\affiliation{Helmholtz-Zentrum Berlin f\"ur Materialien und Energie, Albert-Einstein-Stra{\ss}e 15, 12489 Berlin, Germany}
\author{\surname{Unai} Atxitia\orcidlink{0000-0002-2871-5644}}%
\affiliation{Instituto de Ciencia de Materiales de Madrid, CSIC, Cantoblanco, 28049 Madrid, Spain}
\author{\surname{Clemens} von Korff Schmising\orcidlink{0000-0003-3159-3489}}%
\affiliation{Max-Born-Institut f\"ur Nichtlineare Optik und Kurzzeitspektroskopie, Max-Born-Stra{\ss}e 2A, 12489 Berlin, Germany}
\author{\surname{Wolfgang} Kuch\orcidlink{0000-0002-5764-4574}}%
\affiliation{Institut f\"ur Experimentalphysik, Freie Universit\"at Berlin, Arnimallee 14, 14195 Berlin, Germany}

\date{\today}

\begin{abstract}
The tremendous interest in the technology and underlying physics of all-optical switching of magnetization brings up the question of how fast the switching can occur and how high the frequency of writing the data with ultrafast laser pulses can be. To answer this question,
we excited a GdFe ferrimagnetic alloy, the magnetization of which can be reversed by single laser pulses, a phenomenon known as toggle switching, by two pulses with a certain time delay in between.  Using photoemission electron microscopy and Kerr microscopy for magnetic domain imaging, we 
explore the effects of varying fluences of the first and second pulse as well as the time delay between the two pulses.  
Our results show that when the fluence of the first pulse is adjusted just above the threshold of single-pulse switching, a second pulse with about  \SI{60}{\percent} of the fluence of the first pulse, arriving only \SI{3}{\pico\second}
  later, switches the magnetization back. This reswitching persists up to about \SI{40}{\pico\second} pulse separation. We interpret the latter as the time required for the sample to cool down and remagnetize after the first pulse. For shorter time delays below about \SI{2}{\pico\second}, no re-switching occurs. However, the effect of the two pulses adds up, enabling switching for fluences of both pulses below the threshold for single-pulse switching.  Atomistic spin dynamics simulations are used to model the experimental data, successfully confirming our results.
\end{abstract}
\maketitle

\section{Introduction}
Ultrafast all-optical switching (AOS) of magnetization has garnered significant research due to its importance in both fundamental understanding and potential technological advancements \cite{kimel_fundamentals_2020,kirilyuk_ultrafast_2010}. It offers a promising solution for developing faster and more energy-efficient magnetic data storage devices. AOS was first observed as helicity-dependent
 switching in the GdFeCo ferrimagnetic alloy \cite{Stanciu_PhysRevLett.99.047601}, where right and left circularly polarized laser pulses changed the direction of the
 magnetization in opposite directions without the need for an external magnetic field. Some years later, it was observed that the underlying mechanism is a defined switching with linearly polarized laser pulses, referred to as
 helicity-independent switching or thermal switching \cite{radu_transient_2011,ostler_ultrafast_2012} and has been termed ''toggle switching''.
Toggle switching is not limited to GdFeCo alloys but occurs in a broad range of ferrimagnetic materials, for example in Mn\textsubscript{2}Ru\textsubscript{0.9}Ga Heusler alloy, a rare-earth-free ferrimagnet  \cite{ banerjee_single_2020}. 
The rapid growth of the number of materials showing AOS, alongside extensive research to enhance its understanding, has raised questions regarding the maximum frequency 
at which the magnetization can be re-switched based on AOS. To investigate this, experiments using two single pulses with an adjusted time delay (t\textsubscript{d}) between them are conducted.
In 2021, Wang and colleagues observed that the  Gd\textsubscript{27}Fe\textsubscript{64}Co\textsubscript{9} alloy with magnetic compensation temperature of  T\textsubscript{M} = \SI{470}{\kelvin} requires a minimum delay of \SI{300}{\pico\second} of the second pulse to switch back \cite{wang_dual-shot_2021}. They pointed out that this is the minimum time needed for the alloy to cool down to \SI{470}{\kelvin}, the highest temperature
 at which the sample can switch. The time delay between two pulses should be long
 enough for the sample to reach this temperature, after which the second pulse can induce the switch back. The researchers suggested that altering the composition of the sample and improving the heat diffusion could potentially increase the switching frequency \cite{wang_dual-shot_2021}.
In 2022, Steinbach et al. examined the idea of optimizing heat diffusion using different substrates, such as amorphous glass, diamond, and silicon, which possess different heat conductivities \cite{steinbach_accelerating_2022}. Although they emphasized different fluence thresholds for single-pulse toggle switching on each substrate, the GdFe alloy did not re-switch for time delays below \SI{300}{\pico\second}. They reported a time window
 between \SI{300}{}-\SI{500}{\pico\second} for re-switching on diamond and silicon substrates. However, they demonstrated that reliable re-switching is achievable in a GdCo alloy with a pulse-to-pulse separation of \SI{7}{\pico\second}, approaching \SI{}{\tera\hertz} frequency rates for writing and reading data. 
  In this case, the fluence of the second pulse had to be at least 20\% higher than the one of the first pulse.
Double-pulse switching has also been studied in Mn\textsubscript{2}Ru\textsubscript{0.9}Ga Heusler alloys, where re-switching occurs at \SI{12}{\pico\second}, faster than in GdFe and GdFeCo. Banerjee {\it et al.}\ highlighted that the re-switching frequency is related to the spin-lattice relaxation time and the time needed for magnetic damping \cite{banerjee_single_2020}.
Here, we study double-pulse switching in a GdFe thin-film alloy on a Si substrate using the magnetic imaging techniques photoemission electron microscopy (PEEM) and Kerr microscopy. After determining the threshold for single-pulse switching for an individual laser pulse, we systematically study the sample's response to double-pulse excitation, varying the fluences of both pulses and the time delay between them.   
The study is divided into two parts:  First, we apply double pulses with short time delays of less than or equal to \SI{2}{\pico\second} and image the magnetic state of the sample using PEEM. Depending on the initial temperature of the sample and the fluences of the two pulses,
 different response regimes are observed, including switching, no switching, or multi-domain formation. In this time-delay range, the sample does not re-switch. Second, for longer time delays, we use Kerr microscopy and observe that if the fluence of the first pulse is just above the threshold for single-pulse switching, a weaker second pulse in a certain fluence window of about 40--60\% of the first pulse can re-switch the sample at a time delay as short as \SI{3}{\pico\second} at room temperature (RT).  We compare our experimental results to atomistic spin simulations based on the stochastic Landau-Lifshitz-Gilbert equation and a two-temperature model.  The simulations reproduce the experimental observations of suppression of toggle switching of the magnetization by a weaker second pulse with a few ps delay in a certain fluence window and confirm the re-switching of the sublattice magnetizations by the second pulse.

\section{Experiment}

The sample has the structure Al (\SI{3}{})/Gd{\textsubscript{26}Fe{\textsubscript{74} (\SI{10}{})/Pt (\SI{5}{})/ substrate (nominal layer thicknesses are given in \SI{}{\nano\meter}). The Al serves as capping layer to prevent oxidation. The film was 
prepared on a \SI{100}{\nano\meter} thick thermally oxidized Si(001) substate using magnetron sputter deposition (base pressure $< 10^{-8}$ \SI{}{\milli\bar}) from elemental targets at room temperature. The composition and thickness of the GdFe alloy were determined by Rutherford
backscattering spectrometry. The ferrimagnetic GdFe layer has an out-of-plane easy axis of  magnetization with \SI{7}{\milli\tesla} coercivity at RT. To study the AOS of the sample, we use two microscopic methods, Kerr magneto-optical microscopy and PEEM for acquiring static domain images of
the sample after excitation with single or double laser pulses. For PEEM, the setup at the UE49-PGM SPEEM beamline of the BESSY II synchrotron radiation source was used  \cite{kronast_speem_2016},  with the photon energy of the circularly polarized x rays set at \SI{707.1}{\electron\volt} at the Fe $L_3$ edge. A small electromagnet inside the sample holder allows for applying a magnetic field to the sample to saturate the magnetization before applying laser
pulses. The Kerr microscope at the Max Born Institute \cite{steinbach_wide-field_2021} was used. Here, the sample is located near an external electromagnet that allows to saturate its magnetization.

\noindent For exciting the sample, we used linearly $p$-polarized laser pulses with a wavelength of \SI{800}{\nano\meter}  and a pulse length of \SI{100}{\femto\second} for the optical setup at PEEM, and a wavelength of \SI{1030}{\nano\meter} with a pulse length of \SI{250}{\femto\second} for the Kerr microscope.  Both optical setups include a beam splitter and an
optical delay line, allowing to control the time delay between the two pulses. The two pulses are spatially overlapped on the sample. Since we are using two different setups for imaging, the shape of the laser
footprint on the sample and, consequently, the switching area in the sample is different. Both laser profiles follow Gaussian shapes. At the PEEM setup, the laser is focused on the sample at a grazing angle of \SI{16}{\degree} with
respect to the surface and a spot size of about \SI{11}{} $\times$ \SI{30}{\micro\meter}$^2$ (FWHM). In the Kerr microscope setup, the pump laser reaches the sample at a \SI{76}{\degree} angle relative to the surface, with a spot size of \SI{63}{} $\times$ \SI{66}{\micro\meter}$^2$ (FWHM). As a result of the different incidence angles, 
the incident fluence for the switching threshold also varies between the two 
setups. To avoid any confusion, all fluence values are given as absorbed peak fluence of the GdFe layer, at the center of the laser spot. The absorption of the light in each layer of the sample is calculated and can be found in supplementary Fig.\ S1 \cite{supplement}.
Experimentally, the threshold values of the absorbed fluence determined in that way differ significantly in the two setups, being (at room temperature) \SI{3.4}{\milli\joule\per\centi\meter\textsuperscript{2}} in the PEEM setup and \SI{0.9}{\milli\joule\per\centi\meter\textsuperscript{2}} in the Kerr microscope setup.  A number of possible reasons can be cited for that:  Different actual thresholds for the different pump wavelengths and different temporal pulsewidths, inaccuracies in the determination of the spot sizes, and neglect of interface roughnesses in the calculation of the differential absorption profile.
\section*{Theory}

We performed atomistic spin dynamics (ASD) computer simulations to model the AOS in the sample.
We consider a classical Heisenberg spin Hamiltonian:
\newline
\begin{equation}
\label{Hamiltonian}
     \mathcal{H} = -\sum_{i\neq j\langle ij\rangle} J_{ij} \textbf s_{i} \textbf s_{j}-\sum _{i}d_{i}^{z}( \textbf s_{i}^{z})^{2}
\end{equation}

\noindent Here, $\left |\textbf s_i \right | = 1$ represents the normalized classical spin vector at site $i$. The two sublattices have different and antiparallel atomic magnetic moments $\mu\textsubscript{Fe}$ and $\mu\textsubscript{Gd}$. We consider both intra- ($J\textsubscript{Fe-Fe}$, $J\textsubscript{Gd-Gd} > 0$) and inter-sublattice ($J\textsubscript{Fe-Gd} < 0$) exchange coupling parameters.
It is assumed that the atomic species of the Gd\textsubscript{26}Fe\textsubscript{74} alloy are randomly distributed within the lattice structure. The dynamics of each atomic spin follows the stochastic Landau-Lifshitz-Gilbert (LLG) equation, which can be found in detail with all the parameters used in the supplemental material \cite{supplement}. In our simulations, the electron and lattice energy  dynamics are modeled using the following two-temperature model:

\begin{equation}
\label{twotem_1} 
    C_e(T_e) \frac{\partial{T_{e}}}{\partial{t}} = G_{ep}(T_{ph}-T_{e})+ P_{1}(t) + P_{2}(t)
\end{equation}

\begin{equation}
\label{twotem}
    C_{ph}\frac{\partial{T_{ph}}}{\partial{t}} = G\textsubscript{ep}(T_{e}-T_{ph})+\frac{T_{0}-T_{ph}}{\tau_{d}}
\end{equation}

\noindent where $T_e$, $T_{ph}$, and $T_0$ represent electron, phonon, and equilibrium ambient temperatures, respectively. Here,  $C_e(T_e) = \gamma_eT_e$, with $\gamma_e = 6 \times 10^2$ \SI{}{\joule\meter\textsuperscript{-3}\kelvin\textsuperscript{-2}}.

\noindent To align the electron and lattice temperature dynamics with those calculated by considering the real composition and vertical heat transport along the layers as described in the supplemental material in Fig.\ S8, we slightly increased
 the phonon heat capacity to $C_{ph} = 2.5 \times 10^6 $ \SI{}{\joule\meter\textsuperscript{-3}\kelvin\textsuperscript{-1}} and the electron--phonon coupling to $G_{ep} = 7.0 \times 10^{17}$ \SI{}{\watt\meter\textsuperscript{-3}\kelvin\textsuperscript{-1}}.
The last term in Eq.\ \ref{twotem} corresponds to the phenomenological cooling term that we included in our model. The relaxation time ($\tau_d$) was adjusted to achieve good agreement with the vertical heat flow model across the sample layers. 
For the laser power functions, $P_1(t)$ and $P_2(t)$ in Eq. \ref{twotem_1}, we assumed Gaussian shapes following Ref. \cite{shokr}, to replicate the results of the model with vertical cooling, which are detailed in the supplemental material \cite{supplement}.

\begin{figure*}
    \centering
    \includegraphics[width = 0.8 \linewidth]{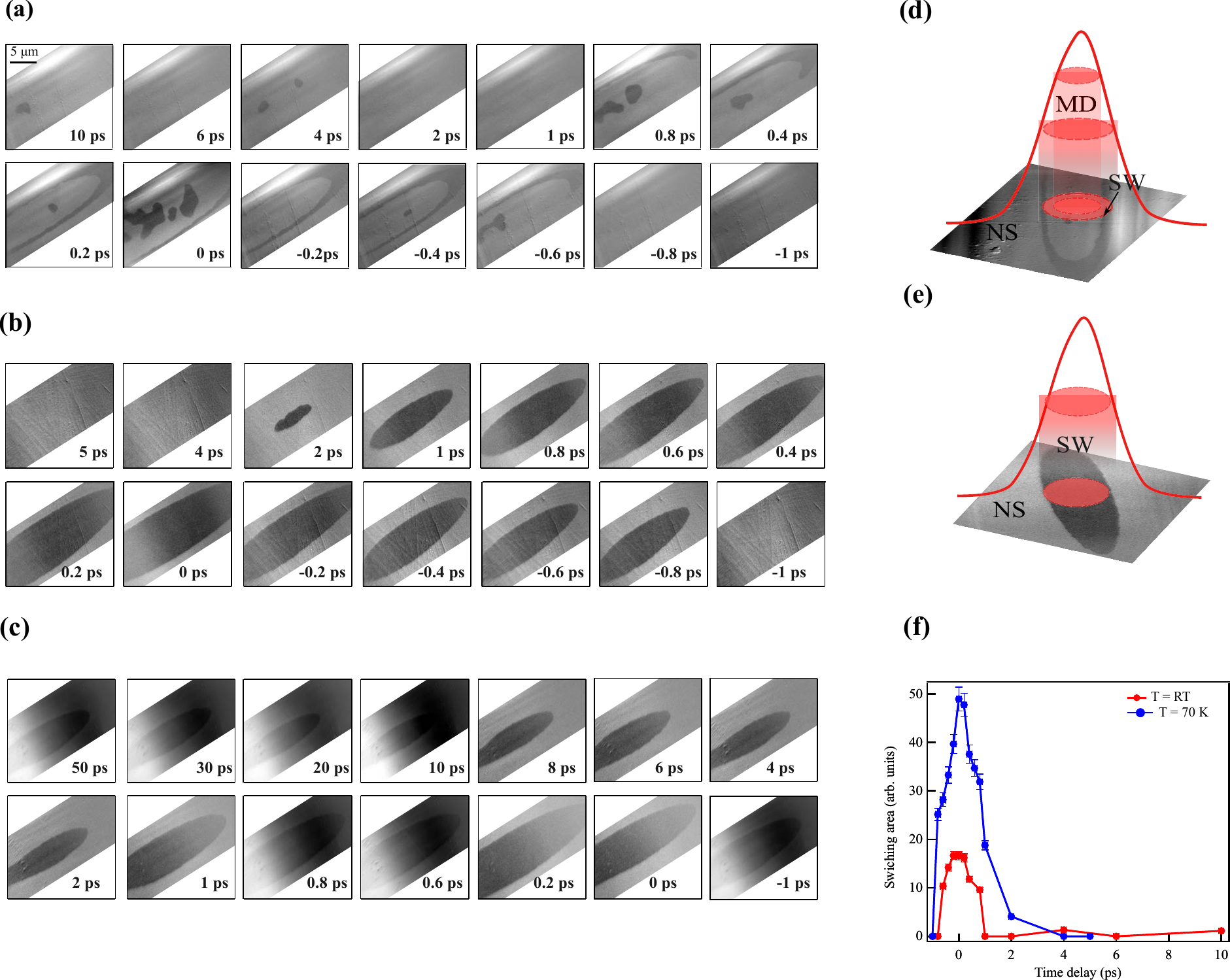}
    \caption{Static XMCD-PEEM images acquired at the Fe $L_3$ edge after double-pulse excitation. Each image presents
     a specific time delay between the two pulses, as written on the top left of each image. Dark and light gray contrast corresponds to opposite
      directions of magnetization. (a) Images measured at room temperature, while the absorbed
       fluence in the GdFe layer is $F_1 =$ \SI{2}{\milli\joule\per\centi\meter\textsuperscript{2}}, $F_2 =$ \SI{2.9}{\milli\joule\per\centi\meter\textsuperscript{2}}. (b) The same experiment performed
        at \SI{70}{\kelvin}, with absorbed fluence in the GdFe layer $F_1$ = \SI{2}{\milli\joule\per\centi\meter\textsuperscript{2}}, $F_2$ = \SI{2.9}{\milli\joule\per\centi\meter\textsuperscript{2}}.
        In both cases, both fluences are below the threshold for magnetic toggle switching, which is  $F_{1th} = F_{2th} =$ \SI{3.4}{\milli\joule\per\centi\meter\textsuperscript{2}}. (c) The same 
        experiment, done at \SI{70}{\kelvin} with absorbed fluence $F_1 =$ \SI{4.7}{\milli\joule\per\centi\meter\textsuperscript{2}},
         $F_2 =$ \SI {0.7}{\milli\joule\per\centi\meter\textsuperscript{2}}, where the fluence of the first pulse is above the threshold of 
         single-pulse toggle switching, which is  $F_{1th}= F_{2th}=$ \SI{3.8}{\milli\joule\per\centi\meter\textsuperscript{2}} at \SI{70}{\kelvin} . (d,e) Sketches of the Gaussian distribution of laser intensity on top of the final magnetic state
          of the sample after double-pulse excitation, show the different regimes of switching at room temperature and \SI{70}{\kelvin}, respectively. (f) The area of the switching 
          after double-pulse excitation for the cases in panels (a) and (b). The field of view in all images  is \SI{20}{}$\times$ \SI{20}{\micro\meter\textsuperscript{2}}.}
    \label{figure1}
\end{figure*}

\section{Results } 

The double-pulse switching behavior of the sample is studied as a function of the fluence for different time delays between the two pulses. Here we categorize our observations in two sections:  first, examining short time delays between pulses of \SI{2}{\pico\second} or less, and second, discussing longer time delays. 
Before each laser excitation, the sample is saturated by an external magnetic field to ensure homogeneous magnetization, meaning the effects imaged after laser excitation result solely from the two laser pulses arriving sequentially with a specific time delay between them. The images represent pixel-by-pixel differences from
the initial saturated state of the sample. Fig.\ref{figure1} (a) shows the switching behavior recorded by PEEM. The
two different contrasts in the images indicate opposite directions of the magnetization; dark contrast represents domains of switched magnetization. The straight bright and dark stripes seen in all images of Fig.\ \ref{figure1} (a) are artifacts caused by
the distribution of x-ray intensity across the field of view. In Fig. \ref{figure1} (a), the sample is at room temperature, and the fluence of both pulses is below the threshold of single-pulse switching, meaning that each pulse alone is insufficient to switch
the sample. However, when  both laser pulses excite the sample, switched areas are observed for time delays between them of less than \SI{1}{\pico\second}. These switched regions appear as irregular dark domains in the center of the laser spot, as well as elliptical patterns
 following the footprint of the laser spot on the sample.  
We can thus distinguish three regions in the sample based on the local fluence of the laser. The first region, located at the center of the laser spot where the fluence is highest, presents a random multi-domain (MD) pattern, probably as a consequence of a
complete demagnetization. The second region, consisting of dark ellipses, corresponds to intermediate fluences located in the gradient of the spatial laser profile and closely follows the shape of the laser spot. In this area, the sample undergoes
deterministic toggle switching (SW). The third region, located at the lowest fluence, outside the ellipses, shows no switching (NS).
The corresponding boundary fluences between these regions vary depending on the time delay between the pulses. For example, the inner line of constant fluence between MD and SW areas at $t_d = \SI{0.2}{\pico\second}$ is approximately
at $F_ 1 = 0.85 F_{1,Max}$, $F_ 2 = 0.83 F_{2,Max}$, where $F_{Max}$ represents the peak fluence, while at $t_d = \SI{0.8}{\pico\second}$, these fluences decreases to approximately $F_ 1 = 0.70 F_{1,Max}$ and $F_ 2 = 0.64 F_{2,Max}$.
\newline Fig. \ref{figure1} (b) depicts the switching behavior of the sample at short time delays at \SI{70}{\kelvin}, which is below the magnetic compensation temperature $T_M$ and the angular momentum compensation temperature $T_A$. As with the RT case, the fluences of both pulses are below the threshold for single-pulse switching.
Similar to the behavior at RT, a switched area appears for time delays shorter than \SI{2}{\pico\second}. However, at \SI{70}{\kelvin}, the sample only shows two distinct regions, the SW region at higher fluences and the NS region at lower fluences.
No multi-domain formation is observed at this temperature.  
The larger size of the SW region compared to Fig.\ref{figure1} (a) indicates that the threshold for switching changes, depending on the initial temperature of the sample, with a lower base temperature reducing the threshold for double-pulse switching. The absence of the MD regime in Fig.\ref{figure1} (b) suggests that the mobility of domain walls is likely higher at RT than at \SI{70}{\kelvin}, facilitating the formation of multi-domain patterns after transient demagnetization of the sample at RT. 
Fig. \ref{figure1} (c) shows the results for the case where the fluence of the first pulse is above the threshold for single-pulse toggle switching while the second pulse remains below this threshold.
The temperature is maintained at \SI{70}{\kelvin}, as in Fig. \ref{figure1} (b). Now a switched area is observed for all time delays caused by the first pulse alone. 
The switched area expands to lower local fluences for delays shorter than \SI{2}{\pico\second}, to regions of the sample that are not switched at larger temporal separations between the two pulses. This behavior is thus consistent with that observed in Fig. 1 (b). The effect of the second pulse obviously enhances the effect of the first pulse at these very short time delays, regardless of whether both pulses are below the threshold or one is above the threshold.
Fig. \ref{figure1} (d) and (e) illustrate the laser profile at the two temperatures to help visualize it. The area of the switching  at RT and   \SI{70}{\kelvin} for the cases in which both pulses are 
below  threshold of the single-pulse switching, is plotted in Fig. \ref{figure1} (f), where it is shown that the switched area 
decreases rapidly at higher time delays. 

\begin{figure*}
    \centering
    \includegraphics[width = 1 \linewidth]{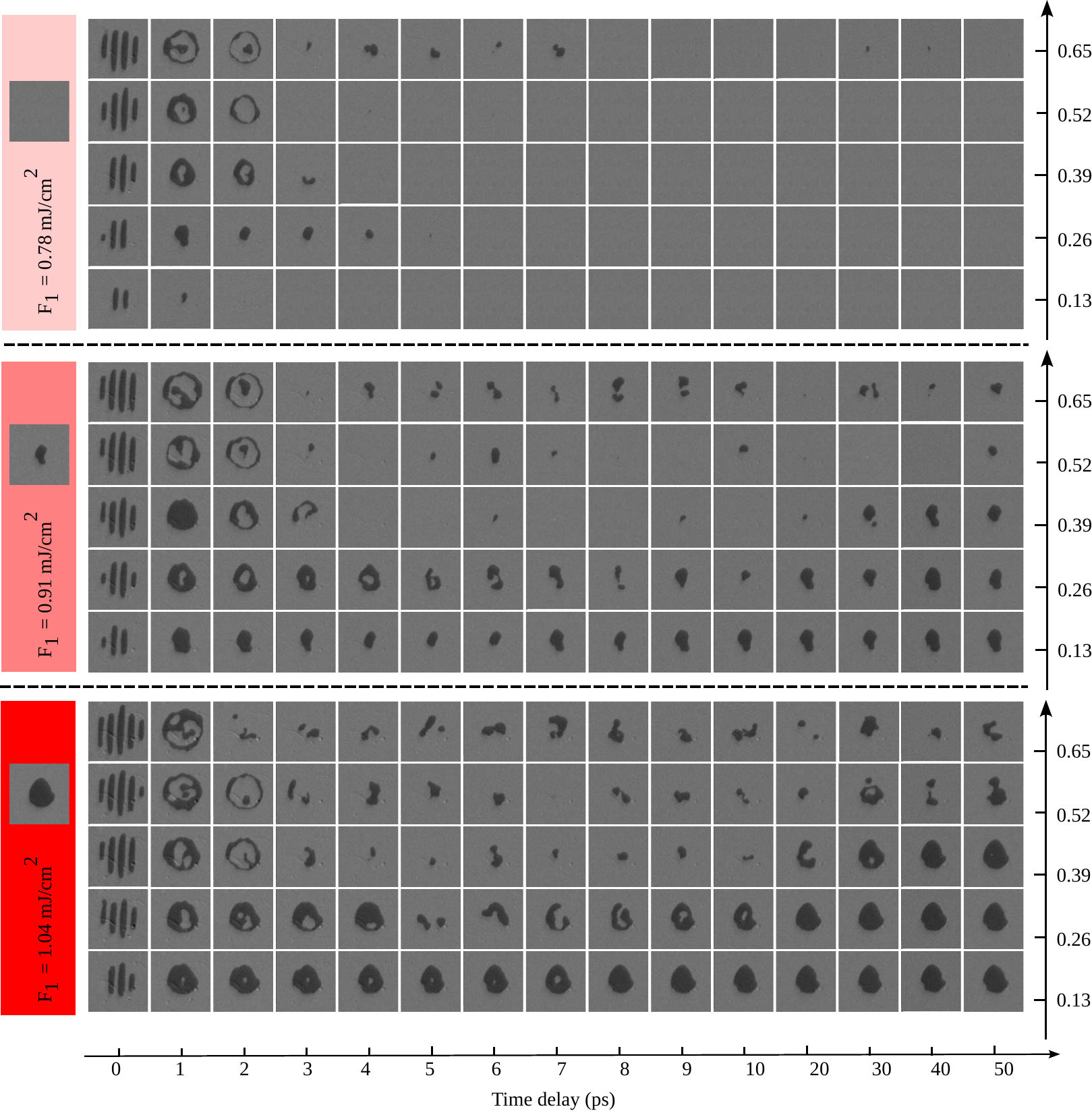}
    \caption{Kerr microscopy images after excitation with two pulses at room temperature. Dark and light gray contrasts correspond to opposite directions of magnetization. The bottom axis shows the time delay between the two pulses. The left column shows the absorbed fluence of the first pulse.
     It also includes an image of single-pulse switching at these fluences. The right axes show the absorbed fluence of the second pulse. The field of view is \SI{50}{}$\times$ \SI{50}{\micro\meter\textsuperscript{2}}.}
    \label{figure2}
\end{figure*}

\noindent In none of these cases does the sample re-switch. To investigate this further, we study the double-pulse switching behavior of the sample using Kerr microscopy at RT, examining different fluences for longer time delays. Fig.\ \ref{figure2} depicts the static state of the sample following double-pulse excitation of the saturated state
 that appears as a light-gray contrast for various fluence combinations and time delays. The figure contains three panels, each corresponding to a different fluence of the first pulse $F_1$, as shown by different colors on the left, along with images
 corresponding to single-pulse switching using only the first pulse for these specific fluences. The first panel corresponds to $F_1 = \SI{0.78}{\milli\joule\per\centi\meter\textsuperscript{2}}$, which is below the threshold for single-pulse 
 toggle switching. The second panel represents $F_1 = \SI{0.91}{\milli\joule\per\centi\meter\textsuperscript{2}}$, just above the threshold, while the third panel presents the results for $F_1 = \SI{1.04}{\milli\joule\per\centi\meter\textsuperscript{2}}$ which
 is well above the threshold but still within the switching window, so a single pulse can excite the sample without demagnetizing it. The bottom axis indicates the time delay between the two pulses. The behavior of single-pulse excitation at different fluences can be found in Fig.\ S2 of the supplemental material \cite{supplement}. 
Both laser pulses are $p$-polarized and spatially overlap on the sample; therefore, at time zero, where both pulses temporally overlap,  an interference pattern appears. In the first panel, with $F_1 = \SI{0.78}{\milli\joule\per\centi\meter\textsuperscript{2}}$ and $t_d = \SI{1}{\pico\second}$, we
 observe only SW region for $F_2 = \SI{0.13}{\milli\joule\per\centi\meter\textsuperscript{2}}$, indicating that, although the sample does not switch with the first pulse alone, adding a second pulse with $F_2 = \SI{0.13}{\milli\joule\per\centi\meter\textsuperscript{2}}$ is enough to switch the
 sample as long as the time delay between the two pulses is \SI{1}{\pico\second} or less, similar to the findings presented before. Increasing $F_2$ to \SI{0.26}{\milli\joule\per\centi\meter\textsuperscript{2}}, allows switching to be observed up to a delay time of \SI{3}{\pico\second}, albeit with a decreasing size of the switched area as the delay increases. From $F_2 = \SI{0.39}{\milli\joule\per\centi\meter\textsuperscript{2}}$, 
a small region of light-gray contrast appears in the center of the switched area, where the fluence is highest, and this region increases further with higher $F_2$ values. This trend is also observed for the other $F_1$ fluences  in Fig.\ \ref{figure2} at \SI{1}{} and \SI{2}{\pico\second} when the sum of $F_1$ and $F_2$ exceeds approximately \SI{1.3}{\milli\joule\per\centi\meter\textsuperscript{2}}.  
As before, this is assigned to a region of demagnetization and domain formation (MD), surrounded by a ring of deterministic toggle switching (SW). This behavior suggests that the sample cannot cool down before the arrival of the second pulse, exhibiting behavior similar to a single pulse with a higher fluence.  Increasing the time delay between the two pulses at $F_1 = \SI{0.78}{\milli\joule\per\centi\meter\textsuperscript{2}}$ brings the sample to the NS state, which means that for this relatively low $F_1$, the effect of the first pulse has already relaxed too much to allow the sample being switched by the second pulse.  However, a nonmonotonic behavior is observed in the time delay at which the NS regime begins as a function of $F_2$.  At $F_2 = \SI{0.52}{\milli\joule\per\centi\meter\textsuperscript{2}}$, already at \SI{3}{\pico\second} no switching is seen anymore, while at both, lower and higher $F_2$, the sample continues to switch at longer time delays.  
\noindent This behavior becomes more pronounced in the center panel, for $F_1 = \SI{0.91}{\milli\joule\per\centi\meter\textsuperscript{2}}$.  At $F_2 = \SI{0.39}{\milli\joule\per\centi\meter\textsuperscript{2}}$, switching is largely suppressed for time delays from \SI{4}{} to \SI{20}{\pico\second}.  This supression range extends even further at the higher second-pulse fluence of $F_2 = \SI{0.52}{\milli\joule\per\centi\meter\textsuperscript{2}}$.  In the bottom panel, for $F_1 = \SI{1.04}{\milli\joule\per\centi\meter\textsuperscript{2}}$, a shrinking of the switched area is evident for time delays between \SI{3}{} and about \SI{10}{\pico\second} at $F_2 = \SI{0.39}{}$ and \SI{0.52}{\milli\joule\per\centi\meter\textsuperscript{2}}. This shrinking indicates that, at fluences slightly below the maximum fluence, corresponding to the center in the Gaussian footprint of the laser pulses, no switching occurs.  
\noindent  This suggests that re-switching of the sample by the second pulse occurs within a specific fluence range of the second pulse, depending on the fluence of the first pulse.  
However, when the fluence of the second pulse is increased excessively, for example to \SI{0.65}{\milli\joule\per\centi\meter\textsuperscript{2}}, 
the sample probably overheats, resulting in the formation of an MD pattern.  
For clarity, the percentage of the switched area relative to the total area in the field of view for all three fluences of the first pulse when $F_2 = $\SI{0.39}{} and $F_2 = \SI{0.52}{\milli\joule\per\centi\meter\textsuperscript{2}}$ presented in Fig. \ref{prob-switching}.
For $F_1 = \SI{0.78}{\milli\joule\per\centi\meter\textsuperscript{2}}$, the switched area drops to zero at longer time delays whereas for $F_1 = \SI{0.91}{\milli\joule\per\centi\meter\textsuperscript{2}}$, the switched area of the sample drops to almost zero for  time delays between \SI{4}{\pico\second} and \SI{20}{\pico\second}.
 For $F_2 = \SI{0.52}{\milli\joule\per\centi\meter\textsuperscript{2}}$, this range extends to \SI{40}{\pico\second}.

\begin{figure*}
    \centering
    \includegraphics[width=0.8\textwidth]{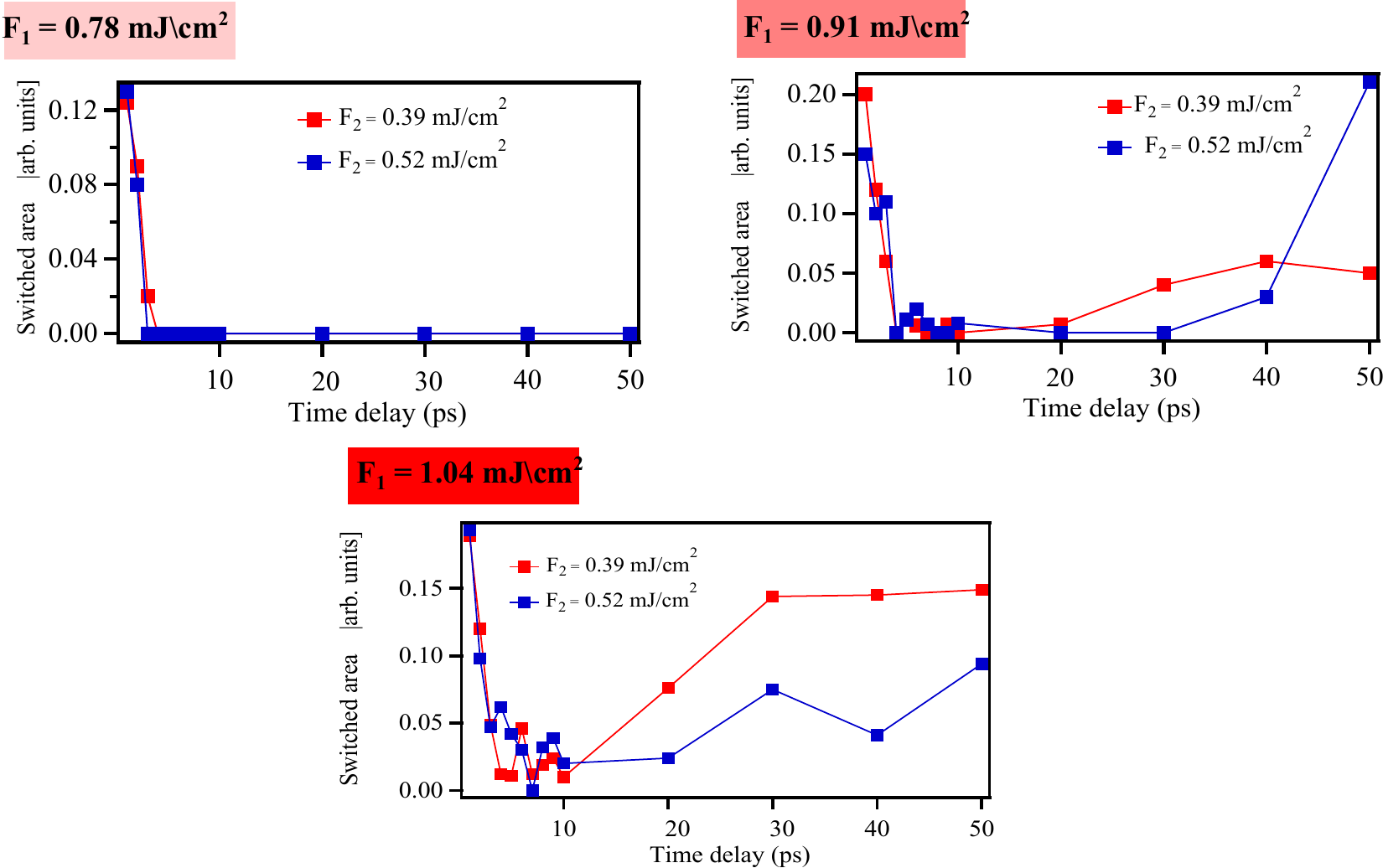}
    \caption{ The probability of the switching as a function of time delay for different fluences of the first pulse.
    (a)  $F_1 = \SI{0.78}{\milli\joule\per\centi\meter\textsuperscript{2}}$, which is below the threshold for single-pulse switching. Red and blue color  are for different fluences of the second pulse. The data show that the
    switching probability  is zero at long time delays.  (b) $F_1 = \SI{0.91}{\milli\joule\per\centi\meter\textsuperscript{2}}$, just above the threshold for single-pulse switching. When $F_2 = \SI{0.39}{\milli\joule\per\centi\meter\textsuperscript{2}}$  (red color),
     re-switching continues until \SI{20}{\pico\second}, while at $F_2 = \SI{0.52}{\milli\joule\per\centi\meter\textsuperscript{2}}$ (blue), re-switching lasts until  \SI{40}{\pico\second}.
  (c)  The trend of reducing the switching probability for $F_1 = \SI{1.04}{\milli\joule\per\centi\meter\textsuperscript{2}}$ and $F_2 = \SI{0.39}{\milli\joule\per\centi\meter\textsuperscript{2}}$ goes on
 until \SI{20}{\pico\second}, while for $F_2 = \SI{0.52}{\milli\joule\per\centi\meter\textsuperscript{2}}$, it lasts until \SI{40}{\pico\second}. }  
    \label{prob-switching}
\end{figure*}

\noindent To study the occurrence of the re-switching more precisely, we focus on $F_1 = \SI{0.91}{\milli\joule\per\centi\meter\textsuperscript{2}}$ while changing $F_2$ in smaller steps. The results are shown in Fig.\ \ref{smallersteps}, where the left axis displays the fluence
 of the second pulse, ranging from \SI{0.13}{} to \SI{0.52}{\milli\joule\per\centi\meter\textsuperscript{2}}, and the bottom axis shows the time delay between the two pulses. As the fluence of the second pulse is
  increased at $t_d = \SI{4}{\pico\second}$, an MD region first develops at the center of the pulse spots, and then, at $F_2 = \SI{0.36}{\milli\joule\per\centi\meter\textsuperscript{2}}$, re-switching begins as indicated by a
   yellow frame. At t\textsubscript{d}= \SI{8}{\pico\second}, re-switching occurs at a lower fluence of $F_2 = \SI{0.31}{\milli\joule\per\centi\meter\textsuperscript{2}}$, which is the minimum fluence of the second pulse required to completely re-switch the sample at this $F_1$.  

\begin{figure} 
    \centering
    \includegraphics[width = 1 \linewidth]{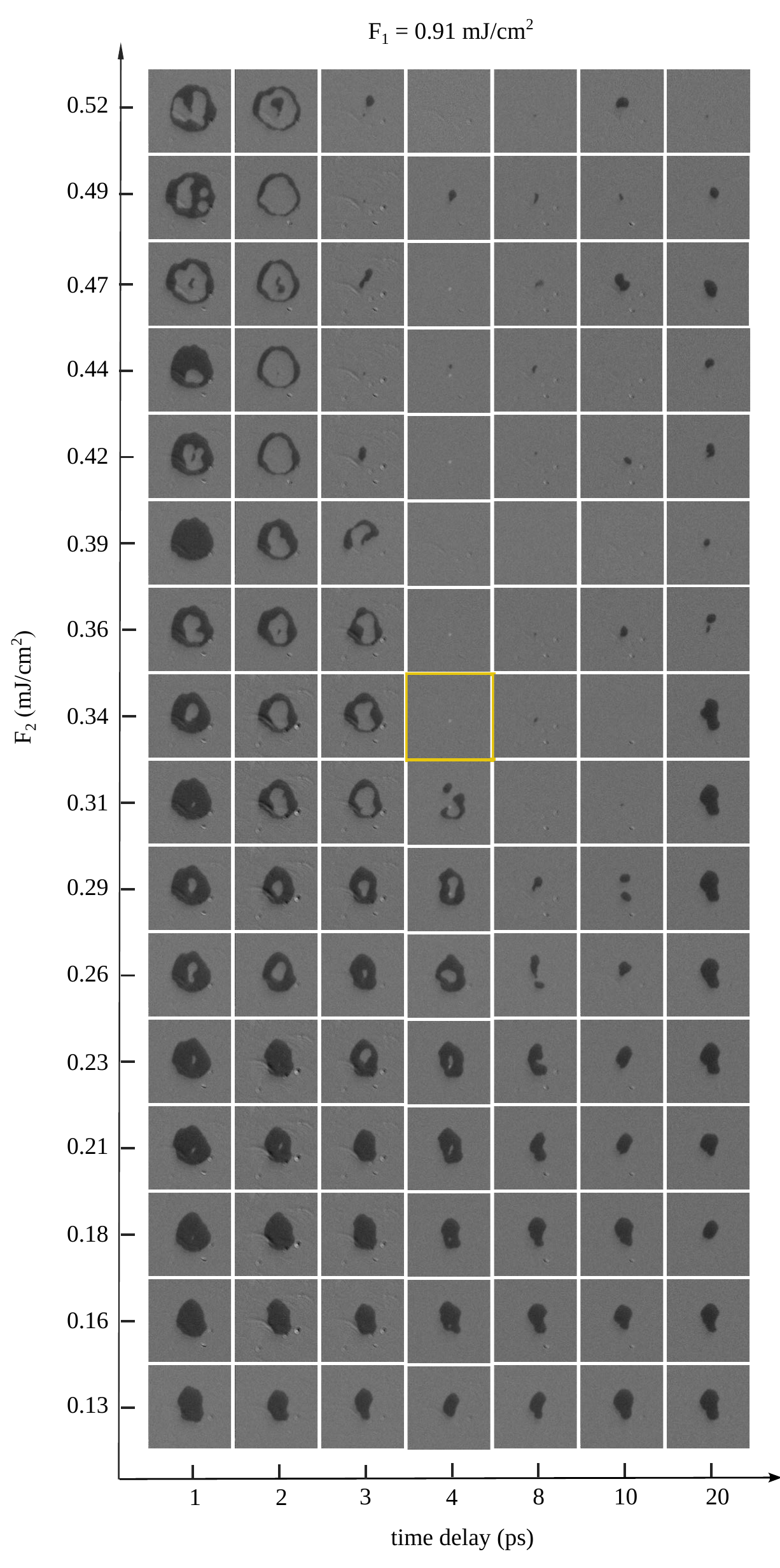}
    \caption{The magnetic state of the sample after double-pulse excitation when the absorbed fluence of the first pulse is \SI{0.91}{\milli\joule\per\centi\meter\textsuperscript{2}},  just above the threshold for all-optical toggle switching. The left axis shows the absorbed fluence of the second pulse increasing in small steps 
     below the threshold for single-pulse switching. The bottom axis shows the time delay. The images are recorded at room temperature and 
     the field of view is \SI{50}{}$\times$ \SI{50}{\micro\meter\textsuperscript{2}}.}
    \label{smallersteps}
\end{figure}

\begin{figure} 
    \centering
    \includegraphics[width=0.5\textwidth]{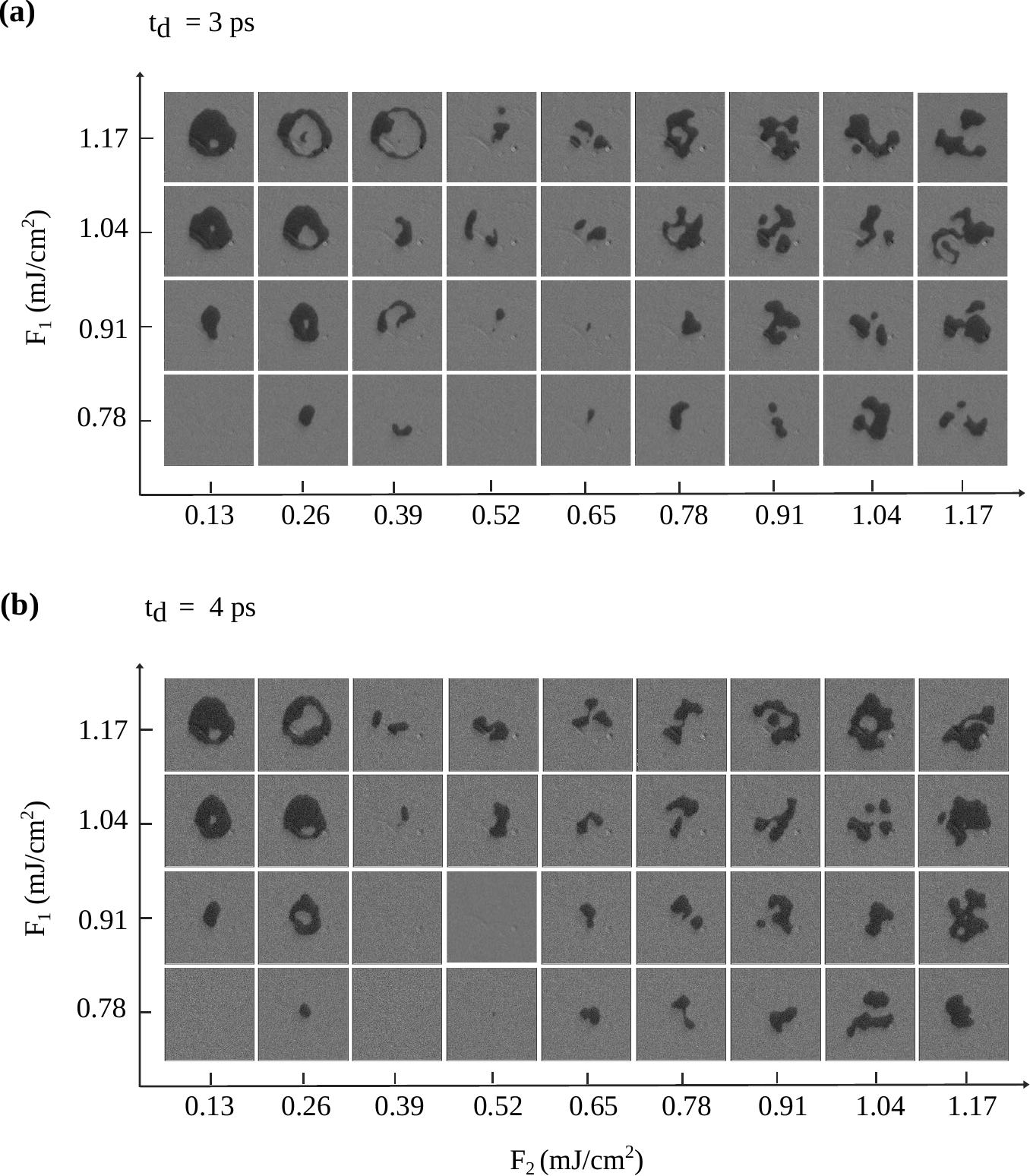}
    \caption{Kerr microscopy images after double-pulse excitation at room temperature at fixed time delay between the pulses of \SI{3}{} and \SI{4}{\pico\second} in (a) and (b), respectively. The bottom axis is the absorbed fluence of the second pulse, and the
     left axis is the absorbed fluence of the first pulse. The field of view is \SI{50}{} $\times$ \SI{50}{\micro\meter\textsuperscript{2}}.}
    \label{figure4}
\end{figure}
\noindent Since re-switching begins at $t_d =  \SI{3}{\pico\second}$ and occurs in a broader range of $F_2$ fluences at $t_d = \SI{4}{\pico\second}$, we looked at the magnetization in more detail at these two time delays.
In Fig.\ \ref{figure4}, the left axis displays the fluence of the first pulse, while the bottom axis shows the fluence of the second pulse for $t_d = \SI{3}{\pico\second}$ (a) and \SI{4}{\pico\second} (b).
In the bottom row of Fig.\ \ref{figure4} (a), where the fluence of the first pulse is below the switching threshold, switching after the second pulse occurs for $F_2 = \SI{0.26}{}$ and $\SI{0.39}{\milli\joule\per\centi\meter\textsuperscript{2}}$, but not for $F_2 = \SI{0.52}{\milli\joule\per\centi\meter\textsuperscript{2}}$.
  For $F_1 = \SI{0.91}{\milli\joule\per\centi\meter\textsuperscript{2}}$, just above the switching threshold, a second pulse with \SI{0.65}{\milli\joule\per\centi\meter\textsuperscript{2}} can re-switch
 the sample, but higher $F_2$ values lead to the formation of MD  patterns.  However, when $F_1 = \SI{1.04}{\milli\joule\per\centi\meter\textsuperscript{2}}$, which is well above the single-pulse switching
  threshold, the same $F_2 = \SI{0.65}{\milli\joule\per\centi\meter\textsuperscript{2}}$ results in MD formation, preventing re-switching.
  Therefore, a certain ratio of $F_2/F_1$  is needed for the sample to successfully re-switch. 
\begin{figure} 
    \centering
    \includegraphics[width=0.5\textwidth]{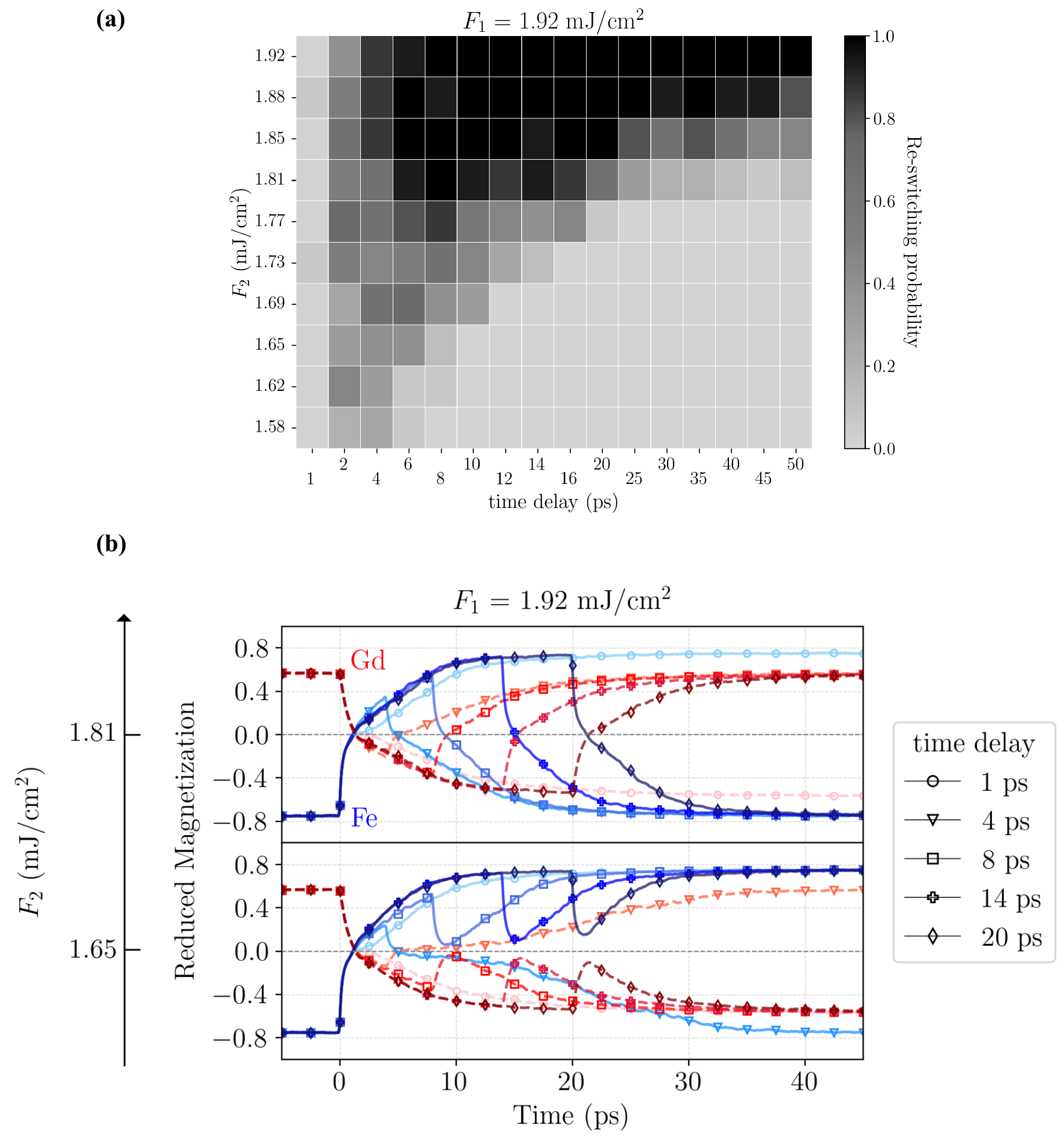}
    \caption{ (a) The calculated probability of re-switching the magnetization after double-pulse excitation. The fluence of the first pulse is constant and corresponds to the threshold for
     single-pulse switching, which is \SI{1.92}{\milli\joule\per\centi\meter\textsuperscript{2}}.  White and black colors in panel (a) corrrespond to no re-switching and re-switching, respectively, while intermediate shades represent the probability of re-switching, as shown in the legend. (b) The magnetization of each sub-lattice, normalized to one for $T = 0$, as a function of time for two selected fluences of the second
      pulse, $F_2 = 1.81$ and \SI{1.65}{\milli\joule\per\centi\meter\textsuperscript{2}}, as shown on the left axis.  The Fe magnetization is shown in blue colors and
       the one of Gd in red. The time delay between the two pulses is indicated by different symbols in the legend.}
    \label{figure5}
\end{figure}
\section{Discussion}

To better understand the mechanism behind double-pulse switching and re-switching, we carried out atomistic spin dynamic (ASD) simulations \cite{PhysRevLett.129.037203, PhysRevB.106.134414,jakobs_exchange-enhancement_2022,jakobs_universal_2022,jakobs_unifying_2021}.
 The re-switching probability from these simulations is shown in Fig.\ \ref{figure5} for a fixed fluence $F_1$ of the first pulse. Black and white colors indicate the probability of re-switching the magnetization by the second pulse, obtained from 15 individual simulations, each corresponding to the outcome of an individual simulation using the LLG equation with the Hamiltonian given in Eq.\ref{Hamiltonian}.  The fluence of the first pulse, $F_1$, is set just above the single-pulse switching threshold.
  In the simulations, this threshold is \SI{1.92}{\milli\joule\per\centi\meter\textsuperscript{2}}, which is roughly twice the experimental value.
 The outcome of individual simulations with combinations of $F_1$, $F_2$, and $t_d$ close to the switching threshold exhibits significant variability due to the nature of  the stochastic Landau-Lifshitz-Gilbert (LLG) equation. 
  The color gradient between black and white in the plot indicates the re-switching probability from 15 individual simulations. In the figure, the fluence of the second pulse increases along the vertical axis while the delay time progresses along the horizontal axis.  This allows for comparison with 
Fig.  \ref{figure2} or Fig. \ref{smallersteps} from the experiment, where one has to keep in mind that the simulations do not include dipolar interactions and thus will not reproduce multi-domain formation. The simulations qualitatively reproduce the same behavior observed experimentally.
at $t_d = \SI{1}{\pico\second}$, the sample switches, but beginning at $ t_d$= \SI{2}{} or \SI{4}{\pico\second}, re-switching occurs for $F_2$ values exceeding a certain threshold, around \SI{1.77}{\milli\joule\per\centi\meter\textsuperscript{2}}.
This consistency between simulations and experiments helps confirm the key role of the time delay and fluence in controlling the switching dynamics, with stochastic effects playing a significant role in determining the exact behavior at specific parameter combinations.
The re-switching dynamics become increasingly complex as the time delay increases.  The maximum delay time after which the presence of the second pulse can hinder the sample from being switched increases with increasing $F_2$, as in the experiment. 
Figure \ref{figure5} (b) shows the simulated time traces of the Gd and Fe sublattice magnetizations for ten distinct combinations of $F_2$ and $t_d$ from panel (a). After the first laser pulse excites the sample,  its energy raises the
 electron temperature $T_e$, which then dissipates to the phonon system via electron-phonon coupling, establishing an equilibrium between the electron and phonon temperatures within \SI{2}{\pico\second}.
The maximum electron temperature reached depends on the fluences of the laser pulses, as elaborated in the supplemental material  \cite{supplement}. 
Previous theoretical work has shown that a certain amount of reversed magnetization of both sublattices, about 20\%, at the time of the second pulse is necessary to induce re-switching \cite{Atxitia2018}.
From the experiment conducted by Radu {\it et al.} \cite{radu_transient_2011} it is known that for single-pulse switching  the Fe sublattice in GdFeCo demagnetizes within \SI{300}{\femto\second}, while the Gd sublattice takes longer, approximately \SI{1.5}{\pico\second}.
Another work checked the XMCD as well as the calculation for  GdFeCo alloy and reported that Fe demagnetizes in \SI{0.4}{\pico\second} and Gd in \SI{1.2}{\pico\second} \cite{radu_ultrafast_2015}.

  \begin{figure}
    \centering
    \includegraphics[width=0.5\textwidth]{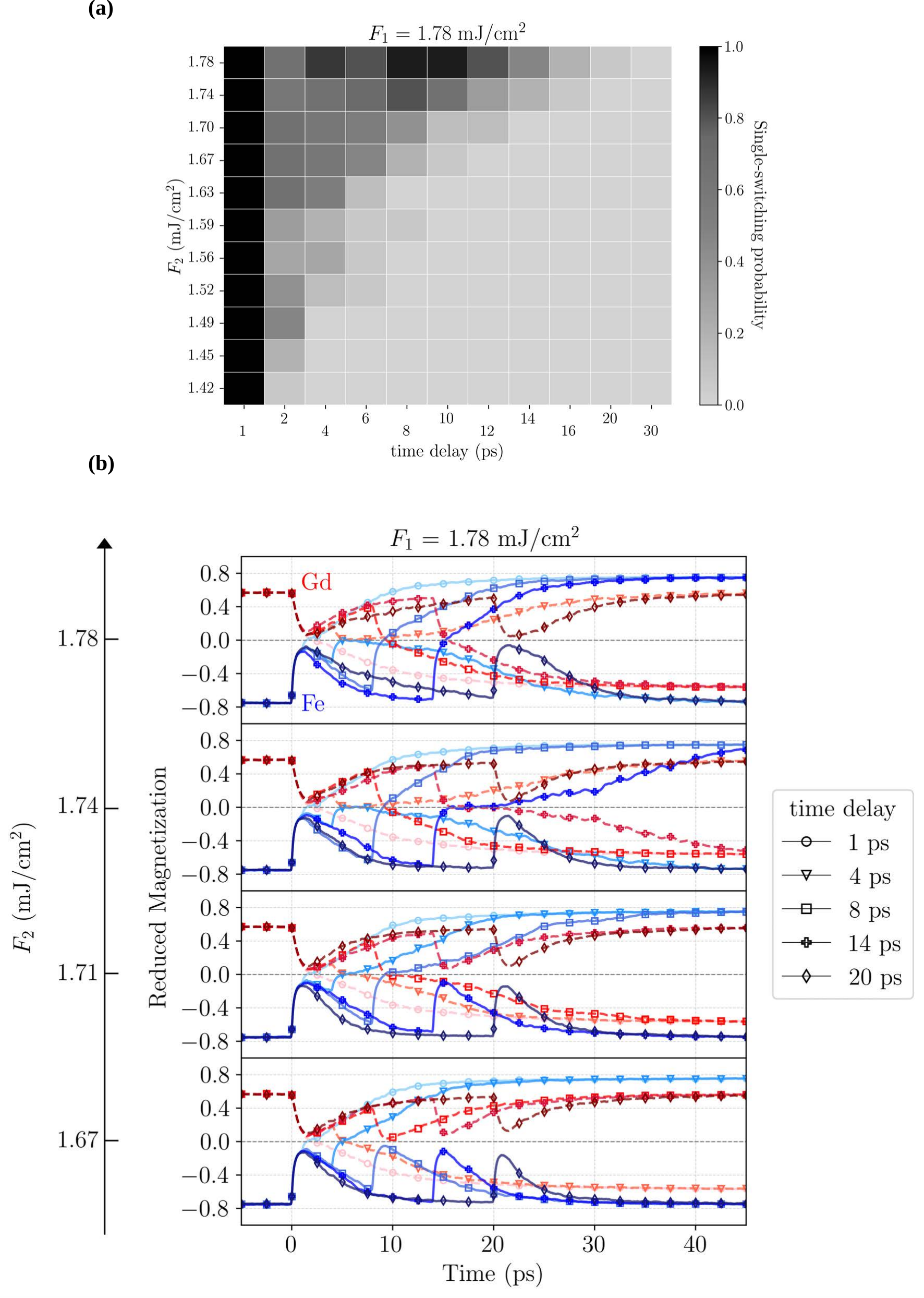}
    \caption{ (a) The probability for switching as a function of the fluence of the second pulse for different time delays. The fluence of the first pulse is constant at $F_1 = \SI{1.78}{\milli\joule\per\centi\meter\textsuperscript{2}}$, which is below the threshold of single-pulse switching.
    Black and white colors correspond to switching and no switching, respectively. (b) The magnetization of each sub-lattice, normalized to one for $T = 0$, as a function of time for four selected fluences of the second
      pulse,  $F_2 = 1.78, 1.74, 1.71 $ and \SI{1.67}{\milli\joule\per\centi\meter\textsuperscript{2}}. 
      The Fe magnetization is shown in blue and the one of Gd in red. The time delay between the two pulses is indicated by different symbols in the legend.}
    \label{figure6}
\end{figure}

Therefore, if the second pulse is applied at $t_d = \SI{1}{\pico\second}$ before the electron and phonon systems reach equilibrium, the sublattices are still in the course of demagnetization (as shown in Fig.\ \ref{figure5} (b)). 
At this point, the magnetization of both sublattices is close to zero and the second pulse further demagnetizes the system without reversing the switching process, since the magnetization has not yet recovered sufficiently for the second pulse to induce re-switching. 
In contrast, for $t_d = \SI{4}{\pico\second}$, both sublattice magnetizations have already crossed zero, making it possible for the second pulse to reverse the magnetization and induce re-switching, as shown in Fig.\ \ref{figure5} (b). 
The longer the delay between the two pulses, the more the magnetization of the sublattices stabilizes in opposite directions as the sample cools further. The more the sample has cooled, the higher fluences 
$ F_2$  are necessary to initiate the re-switching. This is in agreement with the experimental observation that when the first pulse is set at the switching threshold, increasing the fluence of the second pulse results in re-switching being
observed over a larger time-delay window. In the examples selected for Fig.\ \ref{figure5} (b), both at $t_d = \SI{8}{\pico\second}$ and $t_d = \SI{14}{\pico\second}$, a second pulse of \SI{1.81}{\milli\joule\per\centi\meter\textsuperscript{2}} is able to re-switch the sample, while
\SI{1.65}{\milli\joule\per\centi\meter\textsuperscript{2}} is insufficient for re-switching.
According to the experimental result of Ref.\ \cite{radu_transient_2011}, at \SI{3}{\pico\second} the sublattice magnetizations have already crossed zero and begun to re-magnetize in the opposite direction. Therefore, up to \SI{3}{\pico\second}, there is enough time for $T_e$ to cool down, and if the second pulse has enough energy, it can re-switch both sub-lattices.
This has also been recently found in calculations of double-pulse switching in GdFe \cite{Liu2024}. 
The effect of the second pulse also depends on the amount of sublattice magnetization reversed by the first pulse. Shortly after crossing zero, when only a small amount of reversed sublattice magnetization is present, a stronger second pulse is needed to pull the magnetizations back across zero (see, for example, Fig.\ S11 of the supplemental material at 4 ps time delay). 
This is observed in Fig.\ \ref{smallersteps}, where the minimum time required for reswitching increases when decreasing the fluence of the second pulse down to 0.31 mJ/cm$^2$, leading to a narrower delay-time window for reswitching at lower $F_2$.
The reason for this could be that at the higher temperatures being present at shorter delays, the demagnetisation dynamics of the Gd and Fe sublattices become more similar \cite{PhysRevB.89.224421}, making it harder for the second pulse to toggle the magnetization.

At even longer time delays, the sample only toggle-switches, and the second pulse does not have any effect on re-switching or demagnetizing the sample as long as its fluence is below the single-pulse switching threshold.
This behavior is observed in both the experimental data and the simulation results. It is also discussed in the theory work of Ref.\ \cite{liu_minimum_2023} as a possible scenario for double-pulse switching. In this case, the first pulse switches the sample, and by the time the second pulse adds energy, the sample has already cooled down, making it insufficient for either re-switching or demagnetizing.
 A study of Fe/Gd bilayers highlights the ultrafast demagnetization dynamics of both sublattices, with Fe demagnetizing in \SI{0.7}{\pico\second} and Gd in \SI{3.4}{\pico\second} at room temperature (RT) \cite{Dominic_2024}. 
The authors of that study attribute the demagnetization to magnons and argue that the thickness of the individual layers plays a crucial role, as magnons must traverse the layers to transfer magnetic moments to layers farther from the interface \cite{Dominic_2024}, wherein
in GdFeCo alloys, Fe and Gd atoms are separated by just a few {\AA}ngstroms; the transfer of magnetic moments is faster, leading to even shorter demagnetization times.
Steinbach et al. reported re-switching in Gd\textsubscript{24}Fe\textsubscript{76} (\SI{20}{\nano\meter}) on a Si substrate, with a calculated $T_c$ of \SI{515}{\kelvin} occurring approximately at \SI{500}{\pico\second} \cite{steinbach_accelerating_2022} , while for  
GdCo alloys, with a calculated $T_c$ of \SI{570}{\kelvin}, re-switching was observed at \SI{7}{\pico\second} for fluences of the second pulse $> 1.2$ times the one of the first pulse. The different times were discussed as a result of the higher exchange interactions within the transition metal sublattice when replacing Fe with Co in the alloy, leading to faster recovery times and re-switching of GdCo \cite{steinbach_accelerating_2022, PhysRevLett.129.037203}.
In the present study, in contrast, for \SI{10}{\nano\meter} Gd\textsubscript{26}Fe\textsubscript{74} on a Si substrate with a calculated $T_c$ of \SI{540}{\kelvin} , we observe re-switching at \SI{4}{\pico\second} if the fluence of the first pulse is tuned to just above the threshold of single-pulse switching and the fluence of the second pulse is on the one hand sufficiently high to reverse the sublattice magnetizations, but on the other hand also sufficiently low to not lead to full demagnetization and MD formation.  This fast re-switching regime may have been overlooked in previous investigations, or may not be present in different samples with different film thicknesses and alloy compositions.
The $T_c$ of our sample is similar to that of GdCo  of Ref.\ \cite{steinbach_accelerating_2022}, which may explain the fast recovery towards the reversed magnetization after the first pulse in our experiments.
Additionally, $T_M$ is likely different between the different samples. 
 It has been shown theoretically that the minimum pulse separation for re-switching depends strongly on the slope of the net magnetization with temperature at the measurement temperature and thus on the compensation temperature, at which this slope becomes zero \cite{Atxitia2018}.  A higher compensation temperature relative to the measurement temperature allows for faster double switching.  Since $T_M$ varies strongly with the alloy composition, this may also explain the different behavior observed in the different studies.

Fig.\ \ref{figure6} (a) displays the simulated switching probability when the first pulse is set below the threshold of single-pulse switching. The vertical axis represents the fluence of the second pulse and the horizontal axis the delay between the two pulses. This can be compared to the experimental results in the first row of Fig.\ \ref{figure2}, for $F_1 = \SI{0.78}{\milli\joule\per\centi\meter\textsuperscript{2}}$.
The simulations show that, for a time delay of \SI{1}{\pico\second}, the energy from both pulses adds up to switch the sample magnetization, as seen in the experiment. This is also partly the case at $t_d = \SI{2}{\pico\second}$, with a higher probability for larger second-pulse fluences. The dynamics of the sublattice magnetizations are presented in Fig.\ \ref{figure6} (b) for four selected values of $F_2$ and five representative delay times.
 For $t_d = \SI{1}{\pico\second}$, the second pulse arrives while the sample is still demagnetizing from the first pulse, adding energy and pushing the sublattice magnetizations across zero, like one stronger pulse.  
For longer time delays, whether switching occurs or not depends on the fluence of the second pulse. Generally, the higher $F_2$, the greater the switching probability at longer $t_d$. This depends on how much the sublattice
 magnetizations have recovered after the first pulse. More recovery requires a higher fluence of the second pulse to induce switching.  
 In the simulation, at $F_1 = \SI{1.78}{\milli\joule\per\centi\meter\textsuperscript{2}}$, the sample switches more reliably at longer time delays (\SI{10}{} or \SI{14}{\pico\second}) than at shorter 
ones (\SI{2}{} or \SI{4}{\pico\second}).  Inspecting Fig.\ \ref{figure6} (b), this can be understood as follows:  the demagnetizing effect of the second pulse at a fixed $F_2$ depends on the transient 
temperature of the sample, and therefore the sublattice magnetizations when the second pulse arrives.  For example, in the simulation shown in the top panel of Fig. \ref{figure6}(b), both pulses have the same fluence. At \SI{14}{\pico\second} after the first pulse, 
the simulated sublattice magnetizations have nearly fully recovered, meaning the second pulse affects the sublattice magnetization in a manner similar to the first, potentially leading to switching. However, if the second pulse arrives while the sublattice magnetizations are 
still reduced, for example, at  \SI{4}{\pico\second}, the second pulse's effect  on the sublattice magnetizations is diminished which may result in a situation where switching is less likely at shorter delay times than at
 longer ones, as observed in the simulations for $F_2 = \SI{1.74}{\milli\joule\per\centi\meter\textsuperscript{2}}$ at \SI{2}{} and \SI{4}{\pico\second} or at $F_2 = \SI{1.78}{\milli\joule\per\centi\meter\textsuperscript{2}}$ and $t_d = \SI{2}{}, \SI{4}{}$, and $\SI{6}{\pico\second}$. 
In the experiment, there is only one image that shows this effect, namely in Fig.\ \ref{figure2} for $F_1 = \SI{0.78}{\milli\joule\per\centi\meter\textsuperscript{2}}$,
  $F_2 = \SI{0.65}{\milli\joule\per\centi\meter\textsuperscript{2}}$, and $t_d = \SI{3}{\pico\second}$. Since this happens only in one image, it is, however, not clear whether this effect is really observed or simply due to statistical variation.However, the experiment was repeated several times, and the same effect was observed in some of the cases, as shown in supplementary Fig.\ S9.

The re-switching observed experimentally in the top panel of Fig. \ref{figure2} for $F_2 = \SI{0.39}{\milli\joule\per\centi\meter\textsuperscript{2}}$ and \SI{0.52}{\milli\joule\per\centi\meter\textsuperscript{2}} as the absence of switched area for $t_d \ge 3$ respective \SI{8}{\pico\second} could correspond to the reduced switching probability seen in the simulation at $F_2 = \SI{1.74}{\milli\joule\per\centi\meter\textsuperscript{2}}$ and \SI{1.78}{\milli\joule\per\centi\meter\textsuperscript{2}} with time delays of 2 and \SI{4}{\pico\second}.  
However, in the experiment, at these time delays and fluences, significant multi-domain formation already occurs, as shown in Fig.\ \ref{figure4}. Since magnetostatic effects and domain formation are not included in our simulations, the suppression of switching seen at $F_2 = \SI{0.39}{\milli\joule\per\centi\meter\textsuperscript{2}}$ and \SI{0.52}{\milli\joule\per\centi\meter\textsuperscript{2}} in the experiment may not be fully captured by the simulations.

\section{Summary and Conclusions}

In conclusion, we have studied the double-pulse switching of the GdFe alloy using PEEM and Kerr microscopy to understand the effect of the different fluences on the re-switching of the magnetization. By comparing the results to atomistic spin dynamics simulations, we were able to pin down the mechanisms and requirements for all-optical switching by two pulses and for re-switching by a second pulse.
We found that the shortest time between the two pulses at which re-switching of the sample is observed is \SI{4}{\pico\second}, corresponding to \SI{250}{\giga\hertz}.  This is achieved when the fluence of the first pulse is just above the threshold for single-pulse switching and the second pulse has a fluence of about 0.5--0.7 times that of the first pulse.
Re-switching is observed within a certain range of time delays, typically between  \SI{4}{} and \SI{40}{\pico\second}, which depends on the fluence of the second pulse and increases with higher second-pulse fluences.  At these times after the first pulse, the electron and lattice temperature have already equilibrated and the magnetization of each sublattice has moved sufficiently away from zero. However, at shorter temporal distances between the two pulses,
when the electron temperature has not yet equilibrated with the lattice temperature, the effect of the two pulses adds up, leading to switching or multi-domain formation, and thus no back-switching occurs.  
Therefore, to achieve re-switching shortly after a switching event, a fast equilibration of electron and lattice temperatures and efficient heat dissipation to the substrate are required.  An essential ingredient for fast re-switching is to carefully select the fluences of the two pulses, with the fluence of the first pulse slightly above the threshold for all-optical toggle switching and the fluence of the second pulse high enough to trigger another toggle switching event from the not-yet recovered reversed magnetization, but not too high to fully demagnetize the sample.

\begin{acknowledgments}
		This work was supported by the Deutsche Forschungsgemeinschaft via the CRC/TRR 227 ``Ultrafast Spin Dynamics", project-ID: 328545488, projects A02 and A07.  We thank the Helmholtz-Zentrum Berlin for the allocation of synchrotron radiation beamtime.
\end{acknowledgments}

\bibliographystyle{apsrev4-2}

\begin{thebibliography}{7}%
\makeatletter
\providecommand \@ifxundefined [1]{%
 \@ifx{#1\undefined}
}%
\providecommand \@ifnum [1]{%
 \ifnum #1\expandafter \@firstoftwo
 \else \expandafter \@secondoftwo
 \fi
}%
\providecommand \@ifx [1]{%
 \ifx #1\expandafter \@firstoftwo
 \else \expandafter \@secondoftwo
 \fi
}%
\providecommand \natexlab [1]{#1}%
\providecommand \enquote  [1]{``#1''}%
\providecommand \bibnamefont  [1]{#1}%
\providecommand \bibfnamefont [1]{#1}%
\providecommand \citenamefont [1]{#1}%
\providecommand \href@noop [0]{\@secondoftwo}%
\providecommand \href [0]{\begingroup \@sanitize@url \@href}%
\providecommand \@href[1]{\@@startlink{#1}\@@href}%
\providecommand \@@href[1]{\endgroup#1\@@endlink}%
\providecommand \@sanitize@url [0]{\catcode `\\12\catcode `\$12\catcode
  `\&12\catcode `\#12\catcode `\^12\catcode `\_12\catcode `\%12\relax}%
\providecommand \@@startlink[1]{}%
\providecommand \@@endlink[0]{}%
\providecommand \url  [0]{\begingroup\@sanitize@url \@url }%
\providecommand \@url [1]{\endgroup\@href {#1}{\urlprefix }}%
\providecommand \urlprefix  [0]{URL }%
\providecommand \Eprint [0]{\href }%
\providecommand \doibase [0]{https://doi.org/}%
\providecommand \selectlanguage [0]{\@gobble}%
\providecommand \bibinfo  [0]{\@secondoftwo}%
\providecommand \bibfield  [0]{\@secondoftwo}%
\providecommand \translation [1]{[#1]}%
\providecommand \BibitemOpen [0]{}%
\providecommand \bibitemStop [0]{}%
\providecommand \bibitemNoStop [0]{.\EOS\space}%
\providecommand \EOS [0]{\spacefactor3000\relax}%
\providecommand \BibitemShut  [1]{\csname bibitem#1\endcsname}%
\let\auto@bib@innerbib\@empty
\bibitem [{\citenamefont {Ohta}\ and\ \citenamefont
  {Ishida}(1990)}]{ohta_matrix_1990}%
  \BibitemOpen
  \bibfield  {author} {\bibinfo {author} {\bibfnamefont {K.}~\bibnamefont
  {Ohta}}\ and\ \bibinfo {author} {\bibfnamefont {H.}~\bibnamefont {Ishida}},\
  }\bibfield  {title} {\bibinfo {title} {{\color{Gray}Matrix formalism for
  calculation of electric field intensity of light in stratified multilayered
  films}},\ }\href {https://doi.org/10.1364/AO.29.001952} {\bibfield  {journal}
  {\bibinfo  {journal} {Appl. Opt.}\ }\textbf {\bibinfo {volume} {29}},\
  \bibinfo {pages} {1952} (\bibinfo {year} {1990})}\BibitemShut {NoStop}%
\bibitem [{\citenamefont {Chimata}\ \emph {et~al.}(2015)\citenamefont
  {Chimata}, \citenamefont {Isaeva}, \citenamefont {K\'adas}, \citenamefont
  {Bergman}, \citenamefont {Sanyal}, \citenamefont {Mentink}, \citenamefont
  {Katsnelson}, \citenamefont {Rasing}, \citenamefont {Kirilyuk}, \citenamefont
  {Kimel}, \citenamefont {Eriksson},\ and\ \citenamefont
  {Pereiro}}]{PhysRevB.92.094411}%
  \BibitemOpen
  \bibfield  {author} {\bibinfo {author} {\bibfnamefont {R.}~\bibnamefont
  {Chimata}}, \bibinfo {author} {\bibfnamefont {L.}~\bibnamefont {Isaeva}},
  \bibinfo {author} {\bibfnamefont {K.}~\bibnamefont {K\'adas}}, \bibinfo
  {author} {\bibfnamefont {A.}~\bibnamefont {Bergman}}, \bibinfo {author}
  {\bibfnamefont {B.}~\bibnamefont {Sanyal}}, \bibinfo {author} {\bibfnamefont
  {J.~H.}\ \bibnamefont {Mentink}}, \bibinfo {author} {\bibfnamefont {M.~I.}\
  \bibnamefont {Katsnelson}}, \bibinfo {author} {\bibfnamefont
  {T.}~\bibnamefont {Rasing}}, \bibinfo {author} {\bibfnamefont
  {A.}~\bibnamefont {Kirilyuk}}, \bibinfo {author} {\bibfnamefont
  {A.}~\bibnamefont {Kimel}}, \bibinfo {author} {\bibfnamefont
  {O.}~\bibnamefont {Eriksson}},\ and\ \bibinfo {author} {\bibfnamefont
  {M.}~\bibnamefont {Pereiro}},\ }\bibfield  {title} {\bibinfo {title}
  {{\color{Gray}All-thermal switching of amorphous Gd-Fe alloys: Analysis of
  structural properties and magnetization dynamics}},\ }\href
  {https://doi.org/10.1103/PhysRevB.92.094411} {\bibfield  {journal} {\bibinfo
  {journal} {Phys. Rev. B.}\ }\textbf {\bibinfo {volume} {92}},\ \bibinfo
  {pages} {094411} (\bibinfo {year} {2015})}\BibitemShut {NoStop}%
\bibitem [{\citenamefont {Jakobs}\ \emph
  {et~al.}(2021{\natexlab{a}})\citenamefont {Jakobs}, \citenamefont {Ostler},
  \citenamefont {Lambert}, \citenamefont {Yang}, \citenamefont {Salahuddin},
  \citenamefont {Wilson}, \citenamefont {Gorchon}, \citenamefont {Bokor},\ and\
  \citenamefont {Atxitia}}]{PhysRevB.103.104422}%
  \BibitemOpen
  \bibfield  {author} {\bibinfo {author} {\bibfnamefont {F.}~\bibnamefont
  {Jakobs}}, \bibinfo {author} {\bibfnamefont {T.~A.}\ \bibnamefont {Ostler}},
  \bibinfo {author} {\bibfnamefont {C.-H.}\ \bibnamefont {Lambert}}, \bibinfo
  {author} {\bibfnamefont {Y.}~\bibnamefont {Yang}}, \bibinfo {author}
  {\bibfnamefont {S.}~\bibnamefont {Salahuddin}}, \bibinfo {author}
  {\bibfnamefont {R.~B.}\ \bibnamefont {Wilson}}, \bibinfo {author}
  {\bibfnamefont {J.}~\bibnamefont {Gorchon}}, \bibinfo {author} {\bibfnamefont
  {J.}~\bibnamefont {Bokor}},\ and\ \bibinfo {author} {\bibfnamefont
  {U.}~\bibnamefont {Atxitia}},\ }\bibfield  {title} {\bibinfo {title}
  {{\color{Gray}Unifying femtosecond and picosecond single-pulse magnetic
  switching in Gd-Fe-Co}},\ }\href
  {https://doi.org/10.1103/PhysRevB.103.104422} {\bibfield  {journal} {\bibinfo
   {journal} {Phys. Rev. B}\ }\textbf {\bibinfo {volume} {103}},\ \bibinfo
  {pages} {104422} (\bibinfo {year} {2021}{\natexlab{a}})}\BibitemShut
  {NoStop}%
\bibitem [{\citenamefont {Barker}\ \emph {et~al.}(2013)\citenamefont {Barker},
  \citenamefont {Atxitia}, \citenamefont {Ostler}, \citenamefont {Hovorka},
  \citenamefont {Chubykalo-Fesenko},\ and\ \citenamefont
  {Chantrell}}]{barker_two-magnon_2013}%
  \BibitemOpen
  \bibfield  {author} {\bibinfo {author} {\bibfnamefont {J.}~\bibnamefont
  {Barker}}, \bibinfo {author} {\bibfnamefont {U.}~\bibnamefont {Atxitia}},
  \bibinfo {author} {\bibfnamefont {T.~A.}\ \bibnamefont {Ostler}}, \bibinfo
  {author} {\bibfnamefont {O.}~\bibnamefont {Hovorka}}, \bibinfo {author}
  {\bibfnamefont {O.}~\bibnamefont {Chubykalo-Fesenko}},\ and\ \bibinfo
  {author} {\bibfnamefont {R.~W.}\ \bibnamefont {Chantrell}},\ }\bibfield
  {title} {\bibinfo {title} {{\color{Gray}Two-magnon bound state causes
  ultrafast thermally induced magnetisation switching}},\ }\href
  {https://doi.org/10.1038/srep03262} {\bibfield  {journal} {\bibinfo
  {journal} {Sci Rep.}\ }\textbf {\bibinfo {volume} {3}},\ \bibinfo {pages}
  {3262} (\bibinfo {year} {2013})}\BibitemShut {NoStop}%
\bibitem [{\citenamefont {Shokr}\ \emph {et~al.}(2019)\citenamefont {Shokr},
  \citenamefont {Sandig}, \citenamefont {Erkovan}, \citenamefont {Zhang},
  \citenamefont {Bernien}, \citenamefont {{\"U}nal}, \citenamefont {Kronast},
  \citenamefont {Parlak}, \citenamefont {Vogel},\ and\ \citenamefont
  {Kuch}}]{shokr_steering_2019}%
  \BibitemOpen
  \bibfield  {author} {\bibinfo {author} {\bibfnamefont {Y.~A.}\ \bibnamefont
  {Shokr}}, \bibinfo {author} {\bibfnamefont {O.}~\bibnamefont {Sandig}},
  \bibinfo {author} {\bibfnamefont {M.}~\bibnamefont {Erkovan}}, \bibinfo
  {author} {\bibfnamefont {B.}~\bibnamefont {Zhang}}, \bibinfo {author}
  {\bibfnamefont {M.}~\bibnamefont {Bernien}}, \bibinfo {author} {\bibfnamefont
  {A.~A.}\ \bibnamefont {{\"U}nal}}, \bibinfo {author} {\bibfnamefont
  {F.}~\bibnamefont {Kronast}}, \bibinfo {author} {\bibfnamefont
  {U.}~\bibnamefont {Parlak}}, \bibinfo {author} {\bibfnamefont
  {J.}~\bibnamefont {Vogel}},\ and\ \bibinfo {author} {\bibfnamefont
  {W.}~\bibnamefont {Kuch}},\ }\bibfield  {title} {\bibinfo {title}
  {{\color{Gray}Steering of magnetic domain walls by single ultrashort laser
  pulses}},\ }\href {https://doi.org/10.1103/PhysRevB.99.214404} {\bibfield
  {journal} {\bibinfo  {journal} {Phys. Rev. B}\ }\textbf {\bibinfo {volume}
  {99}},\ \bibinfo {pages} {214404} (\bibinfo {year} {2019})}\BibitemShut
  {NoStop}%
\bibitem [{\citenamefont {Jakobs}\ \emph
  {et~al.}(2021{\natexlab{b}})\citenamefont {Jakobs}, \citenamefont {Ostler},
  \citenamefont {Lambert}, \citenamefont {Yang}, \citenamefont {Salahuddin},
  \citenamefont {Wilson}, \citenamefont {Gorchon}, \citenamefont {Bokor},\ and\
  \citenamefont {Atxitia}}]{jakobs_unifying_2021}%
  \BibitemOpen
  \bibfield  {author} {\bibinfo {author} {\bibfnamefont {F.}~\bibnamefont
  {Jakobs}}, \bibinfo {author} {\bibfnamefont {T.~A.}\ \bibnamefont {Ostler}},
  \bibinfo {author} {\bibfnamefont {C.-H.}\ \bibnamefont {Lambert}}, \bibinfo
  {author} {\bibfnamefont {Y.}~\bibnamefont {Yang}}, \bibinfo {author}
  {\bibfnamefont {S.}~\bibnamefont {Salahuddin}}, \bibinfo {author}
  {\bibfnamefont {R.~B.}\ \bibnamefont {Wilson}}, \bibinfo {author}
  {\bibfnamefont {J.}~\bibnamefont {Gorchon}}, \bibinfo {author} {\bibfnamefont
  {J.}~\bibnamefont {Bokor}},\ and\ \bibinfo {author} {\bibfnamefont
  {U.}~\bibnamefont {Atxitia}},\ }\bibfield  {title} {\bibinfo {title}
  {{\color{Gray}Unifying femtosecond and picosecond single-pulse magnetic
  switching in Gd-Fe-Co}},\ }\href
  {https://doi.org/10.1103/PhysRevB.103.104422} {\bibfield  {journal} {\bibinfo
   {journal} {Phys. Rev. B.}\ }\textbf {\bibinfo {volume} {103}},\ \bibinfo
  {pages} {104422} (\bibinfo {year} {2021}{\natexlab{b}})}\BibitemShut
  {NoStop}%
\bibitem [{\citenamefont {Steinbach}\ \emph {et~al.}(2024)\citenamefont
  {Steinbach}, \citenamefont {Atxitia}, \citenamefont {Yao}, \citenamefont
  {Borchert}, \citenamefont {Engel}, \citenamefont {Bencivenga}, \citenamefont
  {Foglia}, \citenamefont {Mincigrucci}, \citenamefont {Pedersoli},
  \citenamefont {De~Angelis}, \citenamefont {Pancaldi}, \citenamefont
  {Fainozzi}, \citenamefont {Pelli~Cresi}, \citenamefont {Paltanin},
  \citenamefont {Capotondi}, \citenamefont {Masciovecchio}, \citenamefont
  {Eisebitt},\ and\ \citenamefont {von
  Korff~Schmising}}]{steinbach_exploring_2024}%
  \BibitemOpen
  \bibfield  {author} {\bibinfo {author} {\bibfnamefont {F.}~\bibnamefont
  {Steinbach}}, \bibinfo {author} {\bibfnamefont {U.}~\bibnamefont {Atxitia}},
  \bibinfo {author} {\bibfnamefont {K.}~\bibnamefont {Yao}}, \bibinfo {author}
  {\bibfnamefont {M.}~\bibnamefont {Borchert}}, \bibinfo {author}
  {\bibfnamefont {D.}~\bibnamefont {Engel}}, \bibinfo {author} {\bibfnamefont
  {F.}~\bibnamefont {Bencivenga}}, \bibinfo {author} {\bibfnamefont
  {L.}~\bibnamefont {Foglia}}, \bibinfo {author} {\bibfnamefont
  {R.}~\bibnamefont {Mincigrucci}}, \bibinfo {author} {\bibfnamefont
  {E.}~\bibnamefont {Pedersoli}}, \bibinfo {author} {\bibfnamefont
  {D.}~\bibnamefont {De~Angelis}}, \bibinfo {author} {\bibfnamefont
  {M.}~\bibnamefont {Pancaldi}}, \bibinfo {author} {\bibfnamefont
  {D.}~\bibnamefont {Fainozzi}}, \bibinfo {author} {\bibfnamefont {J.~S.}\
  \bibnamefont {Pelli~Cresi}}, \bibinfo {author} {\bibfnamefont
  {E.}~\bibnamefont {Paltanin}}, \bibinfo {author} {\bibfnamefont
  {F.}~\bibnamefont {Capotondi}}, \bibinfo {author} {\bibfnamefont
  {C.}~\bibnamefont {Masciovecchio}}, \bibinfo {author} {\bibfnamefont
  {S.}~\bibnamefont {Eisebitt}},\ and\ \bibinfo {author} {\bibfnamefont
  {C.}~\bibnamefont {von Korff~Schmising}},\ }\bibfield  {title} {\bibinfo
  {title} {{\color{Gray}Exploring the Fundamental Spatial Limits of Magnetic
  All-Optical Switching}},\ }\href
  {https://doi.org/10.1021/acs.nanolett.4c00129} {\bibfield  {journal}
  {\bibinfo  {journal} {Nano Lett.}\ }\textbf {\bibinfo {volume} {24}},\
  \bibinfo {pages} {6865} (\bibinfo {year} {2024})}\BibitemShut {NoStop}%
\end{thebibliography}%


\begin{thebibliography}{29}%
\makeatletter
\providecommand \@ifxundefined [1]{%
 \@ifx{#1\undefined}
}%
\providecommand \@ifnum [1]{%
 \ifnum #1\expandafter \@firstoftwo
 \else \expandafter \@secondoftwo
 \fi
}%
\providecommand \@ifx [1]{%
 \ifx #1\expandafter \@firstoftwo
 \else \expandafter \@secondoftwo
 \fi
}%
\providecommand \natexlab [1]{#1}%
\providecommand \enquote  [1]{``#1''}%
\providecommand \bibnamefont  [1]{#1}%
\providecommand \bibfnamefont [1]{#1}%
\providecommand \citenamefont [1]{#1}%
\providecommand \href@noop [0]{\@secondoftwo}%
\providecommand \href [0]{\begingroup \@sanitize@url \@href}%
\providecommand \@href[1]{\@@startlink{#1}\@@href}%
\providecommand \@@href[1]{\endgroup#1\@@endlink}%
\providecommand \@sanitize@url [0]{\catcode `\\12\catcode `\$12\catcode
  `\&12\catcode `\#12\catcode `\^12\catcode `\_12\catcode `\%12\relax}%
\providecommand \@@startlink[1]{}%
\providecommand \@@endlink[0]{}%
\providecommand \url  [0]{\begingroup\@sanitize@url \@url }%
\providecommand \@url [1]{\endgroup\@href {#1}{\urlprefix }}%
\providecommand \urlprefix  [0]{URL }%
\providecommand \Eprint [0]{\href }%
\providecommand \doibase [0]{https://doi.org/}%
\providecommand \selectlanguage [0]{\@gobble}%
\providecommand \bibinfo  [0]{\@secondoftwo}%
\providecommand \bibfield  [0]{\@secondoftwo}%
\providecommand \translation [1]{[#1]}%
\providecommand \BibitemOpen [0]{}%
\providecommand \bibitemStop [0]{}%
\providecommand \bibitemNoStop [0]{.\EOS\space}%
\providecommand \EOS [0]{\spacefactor3000\relax}%
\providecommand \BibitemShut  [1]{\csname bibitem#1\endcsname}%
\let\auto@bib@innerbib\@empty
\bibitem [{\citenamefont {Kimel}\ \emph {et~al.}(2020)\citenamefont {Kimel},
  \citenamefont {Kalashnikova}, \citenamefont {Pogrebna},\ and\ \citenamefont
  {Zvezdin}}]{kimel_fundamentals_2020}%
  \BibitemOpen
  \bibfield  {author} {\bibinfo {author} {\bibfnamefont {A.~V.}\ \bibnamefont
  {Kimel}}, \bibinfo {author} {\bibfnamefont {A.~M.}\ \bibnamefont
  {Kalashnikova}}, \bibinfo {author} {\bibfnamefont {A.}~\bibnamefont
  {Pogrebna}},\ and\ \bibinfo {author} {\bibfnamefont {A.~K.}\ \bibnamefont
  {Zvezdin}},\ }\bibfield  {title} {\bibinfo {title} {{\color{Gray}Fundamentals
  and perspectives of ultrafast photoferroic recording}},\ }\href
  {https://doi.org/10.1016/j.physrep.2020.01.004} {\bibfield  {journal}
  {\bibinfo  {journal} {Phys. Rep.}\ }\textbf {\bibinfo {volume} {852}},\
  \bibinfo {pages} {1} (\bibinfo {year} {2020})}\BibitemShut {NoStop}%
\bibitem [{\citenamefont {Kirilyuk}\ \emph {et~al.}(2010)\citenamefont
  {Kirilyuk}, \citenamefont {Kimel},\ and\ \citenamefont
  {Rasing}}]{kirilyuk_ultrafast_2010}%
  \BibitemOpen
  \bibfield  {author} {\bibinfo {author} {\bibfnamefont {A.}~\bibnamefont
  {Kirilyuk}}, \bibinfo {author} {\bibfnamefont {A.~V.}\ \bibnamefont
  {Kimel}},\ and\ \bibinfo {author} {\bibfnamefont {T.}~\bibnamefont
  {Rasing}},\ }\bibfield  {title} {\bibinfo {title} {{\color{Gray}Ultrafast
  optical manipulation of magnetic order}},\ }\href
  {https://doi.org/10.1103/RevModPhys.82.2731} {\bibfield  {journal} {\bibinfo
  {journal} {Rev. Mod. Phys.}\ }\textbf {\bibinfo {volume} {82}},\ \bibinfo
  {pages} {2731} (\bibinfo {year} {2010})}\BibitemShut {NoStop}%
\bibitem [{\citenamefont {Stanciu}\ \emph {et~al.}(2007)\citenamefont
  {Stanciu}, \citenamefont {Hansteen}, \citenamefont {Kimel}, \citenamefont
  {Kirilyuk}, \citenamefont {Tsukamoto}, \citenamefont {Itoh},\ and\
  \citenamefont {Rasing}}]{Stanciu_PhysRevLett.99.047601}%
  \BibitemOpen
  \bibfield  {author} {\bibinfo {author} {\bibfnamefont {C.~D.}\ \bibnamefont
  {Stanciu}}, \bibinfo {author} {\bibfnamefont {F.}~\bibnamefont {Hansteen}},
  \bibinfo {author} {\bibfnamefont {A.~V.}\ \bibnamefont {Kimel}}, \bibinfo
  {author} {\bibfnamefont {A.}~\bibnamefont {Kirilyuk}}, \bibinfo {author}
  {\bibfnamefont {A.}~\bibnamefont {Tsukamoto}}, \bibinfo {author}
  {\bibfnamefont {A.}~\bibnamefont {Itoh}},\ and\ \bibinfo {author}
  {\bibfnamefont {T.}~\bibnamefont {Rasing}},\ }\bibfield  {title} {\bibinfo
  {title} {{\color{Gray}All-Optical Magnetic Recording with Circularly
  Polarized Light}},\ }\href {https://doi.org/10.1103/PhysRevLett.99.047601}
  {\bibfield  {journal} {\bibinfo  {journal} {Phys. Rev. Lett.}\ }\textbf
  {\bibinfo {volume} {99}},\ \bibinfo {pages} {047601} (\bibinfo {year}
  {2007})}\BibitemShut {NoStop}%
\bibitem [{\citenamefont {Radu}\ \emph {et~al.}(2011)\citenamefont {Radu},
  \citenamefont {Vahaplar}, \citenamefont {Stamm}, \citenamefont {Kachel},
  \citenamefont {Pontius}, \citenamefont {Dürr}, \citenamefont {Ostler},
  \citenamefont {Barker}, \citenamefont {Evans}, \citenamefont {Chantrell},
  \citenamefont {Tsukamoto}, \citenamefont {Itoh}, \citenamefont {Kirilyuk},
  \citenamefont {Rasing},\ and\ \citenamefont {Kimel}}]{radu_transient_2011}%
  \BibitemOpen
  \bibfield  {author} {\bibinfo {author} {\bibfnamefont {I.}~\bibnamefont
  {Radu}}, \bibinfo {author} {\bibfnamefont {K.}~\bibnamefont {Vahaplar}},
  \bibinfo {author} {\bibfnamefont {C.}~\bibnamefont {Stamm}}, \bibinfo
  {author} {\bibfnamefont {T.}~\bibnamefont {Kachel}}, \bibinfo {author}
  {\bibfnamefont {N.}~\bibnamefont {Pontius}}, \bibinfo {author} {\bibfnamefont
  {H.~A.}\ \bibnamefont {Dürr}}, \bibinfo {author} {\bibfnamefont {T.~A.}\
  \bibnamefont {Ostler}}, \bibinfo {author} {\bibfnamefont {J.}~\bibnamefont
  {Barker}}, \bibinfo {author} {\bibfnamefont {R.~F.~L.}\ \bibnamefont
  {Evans}}, \bibinfo {author} {\bibfnamefont {R.~W.}\ \bibnamefont
  {Chantrell}}, \bibinfo {author} {\bibfnamefont {A.}~\bibnamefont
  {Tsukamoto}}, \bibinfo {author} {\bibfnamefont {A.}~\bibnamefont {Itoh}},
  \bibinfo {author} {\bibfnamefont {A.}~\bibnamefont {Kirilyuk}}, \bibinfo
  {author} {\bibfnamefont {T.}~\bibnamefont {Rasing}},\ and\ \bibinfo {author}
  {\bibfnamefont {A.~V.}\ \bibnamefont {Kimel}},\ }\bibfield  {title} {\bibinfo
  {title} {{\color{Gray}Transient ferromagnetic-like state mediating ultrafast
  reversal of antiferromagnetically coupled spins}},\ }\href
  {https://doi.org/10.1038/nature09901} {\bibfield  {journal} {\bibinfo
  {journal} {Nature}\ }\textbf {\bibinfo {volume} {472}},\ \bibinfo {pages}
  {205} (\bibinfo {year} {2011})}\BibitemShut {NoStop}%
\bibitem [{\citenamefont {Ostler}\ \emph {et~al.}(2012)\citenamefont {Ostler},
  \citenamefont {Barker}, \citenamefont {Evans}, \citenamefont {Chantrell},
  \citenamefont {Atxitia}, \citenamefont {Chubykalo-Fesenko}, \citenamefont
  {El~Moussaoui}, \citenamefont {Le~Guyader}, \citenamefont {Mengotti},
  \citenamefont {Heyderman}, \citenamefont {Nolting}, \citenamefont
  {Tsukamoto}, \citenamefont {Itoh}, \citenamefont {Afanasiev}, \citenamefont
  {Ivanov}, \citenamefont {Kalashnikova}, \citenamefont {Vahaplar},
  \citenamefont {Mentink}, \citenamefont {Kirilyuk}, \citenamefont {Rasing},\
  and\ \citenamefont {Kimel}}]{ostler_ultrafast_2012}%
  \BibitemOpen
  \bibfield  {author} {\bibinfo {author} {\bibfnamefont {T.}~\bibnamefont
  {Ostler}}, \bibinfo {author} {\bibfnamefont {J.}~\bibnamefont {Barker}},
  \bibinfo {author} {\bibfnamefont {R.}~\bibnamefont {Evans}}, \bibinfo
  {author} {\bibfnamefont {R.}~\bibnamefont {Chantrell}}, \bibinfo {author}
  {\bibfnamefont {U.}~\bibnamefont {Atxitia}}, \bibinfo {author} {\bibfnamefont
  {O.}~\bibnamefont {Chubykalo-Fesenko}}, \bibinfo {author} {\bibfnamefont
  {S.}~\bibnamefont {El~Moussaoui}}, \bibinfo {author} {\bibfnamefont
  {L.}~\bibnamefont {Le~Guyader}}, \bibinfo {author} {\bibfnamefont
  {E.}~\bibnamefont {Mengotti}}, \bibinfo {author} {\bibfnamefont
  {L.}~\bibnamefont {Heyderman}}, \bibinfo {author} {\bibfnamefont
  {F.}~\bibnamefont {Nolting}}, \bibinfo {author} {\bibfnamefont
  {A.}~\bibnamefont {Tsukamoto}}, \bibinfo {author} {\bibfnamefont
  {A.}~\bibnamefont {Itoh}}, \bibinfo {author} {\bibfnamefont {D.}~\bibnamefont
  {Afanasiev}}, \bibinfo {author} {\bibfnamefont {B.}~\bibnamefont {Ivanov}},
  \bibinfo {author} {\bibfnamefont {A.}~\bibnamefont {Kalashnikova}}, \bibinfo
  {author} {\bibfnamefont {K.}~\bibnamefont {Vahaplar}}, \bibinfo {author}
  {\bibfnamefont {J.}~\bibnamefont {Mentink}}, \bibinfo {author} {\bibfnamefont
  {A.}~\bibnamefont {Kirilyuk}}, \bibinfo {author} {\bibfnamefont
  {T.}~\bibnamefont {Rasing}},\ and\ \bibinfo {author} {\bibfnamefont
  {A.}~\bibnamefont {Kimel}},\ }\bibfield  {title} {\bibinfo {title}
  {{\color{Gray}Ultrafast heating as a sufficient stimulus for magnetization
  reversal in a ferrimagnet}},\ }\href {https://doi.org/10.1038/ncomms1666}
  {\bibfield  {journal} {\bibinfo  {journal} {Nat. Commun.}\ }\textbf {\bibinfo
  {volume} {3}},\ \bibinfo {pages} {666} (\bibinfo {year} {2012})}\BibitemShut
  {NoStop}%
\bibitem [{\citenamefont {Banerjee}\ \emph {et~al.}(2020)\citenamefont
  {Banerjee}, \citenamefont {Teichert}, \citenamefont {Siewierska},
  \citenamefont {Gercsi}, \citenamefont {Atcheson}, \citenamefont {Stamenov},
  \citenamefont {Rode}, \citenamefont {Coey},\ and\ \citenamefont
  {Besbas}}]{banerjee_single_2020}%
  \BibitemOpen
  \bibfield  {author} {\bibinfo {author} {\bibfnamefont {C.}~\bibnamefont
  {Banerjee}}, \bibinfo {author} {\bibfnamefont {N.}~\bibnamefont {Teichert}},
  \bibinfo {author} {\bibfnamefont {K.~E.}\ \bibnamefont {Siewierska}},
  \bibinfo {author} {\bibfnamefont {Z.}~\bibnamefont {Gercsi}}, \bibinfo
  {author} {\bibfnamefont {G.~Y.~P.}\ \bibnamefont {Atcheson}}, \bibinfo
  {author} {\bibfnamefont {P.}~\bibnamefont {Stamenov}}, \bibinfo {author}
  {\bibfnamefont {K.}~\bibnamefont {Rode}}, \bibinfo {author} {\bibfnamefont
  {J.~M.~D.}\ \bibnamefont {Coey}},\ and\ \bibinfo {author} {\bibfnamefont
  {J.}~\bibnamefont {Besbas}},\ }\bibfield  {title} {\bibinfo {title}
  {{\color{Gray}Single pulse all-optical toggle switching of magnetization
  without gadolinium in the ferrimagnet Mn2RuxGa}},\ }\href
  {https://doi.org/10.1038/s41467-020-18340-9} {\bibfield  {journal} {\bibinfo
  {journal} {Nat. Commun.}\ }\textbf {\bibinfo {volume} {11}},\ \bibinfo
  {pages} {4444} (\bibinfo {year} {2020})}\BibitemShut {NoStop}%
\bibitem [{\citenamefont {Wang}\ \emph {et~al.}(2021)\citenamefont {Wang},
  \citenamefont {Wei}, \citenamefont {Feng}, \citenamefont {Cao}, \citenamefont
  {Li}, \citenamefont {Cao}, \citenamefont {Guan}, \citenamefont {Tsukamoto},
  \citenamefont {Kirilyuk}, \citenamefont {Kimel},\ and\ \citenamefont
  {Li}}]{wang_dual-shot_2021}%
  \BibitemOpen
  \bibfield  {author} {\bibinfo {author} {\bibfnamefont {S.}~\bibnamefont
  {Wang}}, \bibinfo {author} {\bibfnamefont {C.}~\bibnamefont {Wei}}, \bibinfo
  {author} {\bibfnamefont {Y.}~\bibnamefont {Feng}}, \bibinfo {author}
  {\bibfnamefont {H.}~\bibnamefont {Cao}}, \bibinfo {author} {\bibfnamefont
  {W.}~\bibnamefont {Li}}, \bibinfo {author} {\bibfnamefont {Y.}~\bibnamefont
  {Cao}}, \bibinfo {author} {\bibfnamefont {B.}~\bibnamefont {Guan}}, \bibinfo
  {author} {\bibfnamefont {A.}~\bibnamefont {Tsukamoto}}, \bibinfo {author}
  {\bibfnamefont {A.}~\bibnamefont {Kirilyuk}}, \bibinfo {author}
  {\bibfnamefont {A.~V.}\ \bibnamefont {Kimel}},\ and\ \bibinfo {author}
  {\bibfnamefont {X.}~\bibnamefont {Li}},\ }\bibfield  {title} {\bibinfo
  {title} {{\color{Gray}Dual-shot dynamics and ultimate frequency of
  all-optical magnetic recording on GdFeCo}},\ }\href
  {https://doi.org/10.1038/s41377-020-00451-z} {\bibfield  {journal} {\bibinfo
  {journal} {Light Sci. Appl.}\ }\textbf {\bibinfo {volume} {10}},\ \bibinfo
  {pages} {8} (\bibinfo {year} {2021})}\BibitemShut {NoStop}%
\bibitem [{\citenamefont {Steinbach}\ \emph {et~al.}(2022)\citenamefont
  {Steinbach}, \citenamefont {Stetzuhn}, \citenamefont {Engel}, \citenamefont
  {Atxitia}, \citenamefont {Von Korff~Schmising},\ and\ \citenamefont
  {Eisebitt}}]{steinbach_accelerating_2022}%
  \BibitemOpen
  \bibfield  {author} {\bibinfo {author} {\bibfnamefont {F.}~\bibnamefont
  {Steinbach}}, \bibinfo {author} {\bibfnamefont {N.}~\bibnamefont {Stetzuhn}},
  \bibinfo {author} {\bibfnamefont {D.}~\bibnamefont {Engel}}, \bibinfo
  {author} {\bibfnamefont {U.}~\bibnamefont {Atxitia}}, \bibinfo {author}
  {\bibfnamefont {C.}~\bibnamefont {Von Korff~Schmising}},\ and\ \bibinfo
  {author} {\bibfnamefont {S.}~\bibnamefont {Eisebitt}},\ }\bibfield  {title}
  {\bibinfo {title} {{\color{Gray}Accelerating double pulse all-optical
  write/erase cycles in metallic ferrimagnets}},\ }\href
  {https://doi.org/10.1063/5.0080351} {\bibfield  {journal} {\bibinfo
  {journal} {Appl. Phys. Lett.}\ }\textbf {\bibinfo {volume} {120}},\ \bibinfo
  {pages} {112406} (\bibinfo {year} {2022})}\BibitemShut {NoStop}%
\bibitem [{\citenamefont {Kronast}\ and\ \citenamefont
  {Valencia~Molina}(2016)}]{kronast_speem_2016}%
  \BibitemOpen
  \bibfield  {author} {\bibinfo {author} {\bibfnamefont {F.}~\bibnamefont
  {Kronast}}\ and\ \bibinfo {author} {\bibfnamefont {S.}~\bibnamefont
  {Valencia~Molina}},\ }\bibfield  {title} {\bibinfo {title}
  {{\color{Gray}SPEEM: The photoemission microscope at the dedicated microfocus
  PGM beamline UE49-PGMa at BESSY II}},\ }\href
  {https://doi.org/10.17815/jlsrf-2-86} {\bibfield  {journal} {\bibinfo
  {journal} {JLSRF.}\ }\textbf {\bibinfo {volume} {2}},\ \bibinfo {pages} {A90}
  (\bibinfo {year} {2016})}\BibitemShut {NoStop}%
\bibitem [{\citenamefont {Steinbach}\ \emph {et~al.}(2021)\citenamefont
  {Steinbach}, \citenamefont {Schick}, \citenamefont {Von Korff~Schmising},
  \citenamefont {Yao}, \citenamefont {Borchert}, \citenamefont {Engel},\ and\
  \citenamefont {Eisebitt}}]{steinbach_wide-field_2021}%
  \BibitemOpen
  \bibfield  {author} {\bibinfo {author} {\bibfnamefont {F.}~\bibnamefont
  {Steinbach}}, \bibinfo {author} {\bibfnamefont {D.}~\bibnamefont {Schick}},
  \bibinfo {author} {\bibfnamefont {C.}~\bibnamefont {Von Korff~Schmising}},
  \bibinfo {author} {\bibfnamefont {K.}~\bibnamefont {Yao}}, \bibinfo {author}
  {\bibfnamefont {M.}~\bibnamefont {Borchert}}, \bibinfo {author}
  {\bibfnamefont {W.~D.}\ \bibnamefont {Engel}},\ and\ \bibinfo {author}
  {\bibfnamefont {S.}~\bibnamefont {Eisebitt}},\ }\bibfield  {title} {\bibinfo
  {title} {{\color{Gray}Wide-field magneto-optical microscope to access
  quantitative magnetization dynamics with femtosecond temporal and
  sub-micrometer spatial resolution}},\ }\href
  {https://doi.org/10.1063/5.0060091} {\bibfield  {journal} {\bibinfo
  {journal} {J. Appl. Phys.}\ }\textbf {\bibinfo {volume} {130}},\ \bibinfo
  {pages} {083905} (\bibinfo {year} {2021})}\BibitemShut {NoStop}%
\bibitem [{sup()}]{supplement}%
  \BibitemOpen
  \href@noop {} {\bibinfo {title} {{\color{Gray}See {S}upplemental {M}aterial
  at [{URL} will be inserted by publisher] for additional details, referencing
  Refs.
  \cite{ohta_matrix_1990,PhysRevB.92.094411,PhysRevB.103.104422,barker_two-magnon_2013,shokr_steering_2019,steinbach_exploring_2024}}}}\BibitemShut
  {NoStop}%
\bibitem [{\citenamefont {Shokr}\ \emph
  {et~al.}(2019{\natexlab{a}})\citenamefont {Shokr}, \citenamefont {Sandig},
  \citenamefont {Erkovan}, \citenamefont {Zhang}, \citenamefont {Bernien},
  \citenamefont {Ünal}, \citenamefont {Kronast}, \citenamefont {Parlak},
  \citenamefont {Vogel},\ and\ \citenamefont {Kuch}}]{shokr}%
  \BibitemOpen
  \bibfield  {author} {\bibinfo {author} {\bibfnamefont {Y.~A.}\ \bibnamefont
  {Shokr}}, \bibinfo {author} {\bibfnamefont {O.}~\bibnamefont {Sandig}},
  \bibinfo {author} {\bibfnamefont {M.}~\bibnamefont {Erkovan}}, \bibinfo
  {author} {\bibfnamefont {B.}~\bibnamefont {Zhang}}, \bibinfo {author}
  {\bibfnamefont {M.}~\bibnamefont {Bernien}}, \bibinfo {author} {\bibfnamefont
  {A.~A.}\ \bibnamefont {Ünal}}, \bibinfo {author} {\bibfnamefont
  {F.}~\bibnamefont {Kronast}}, \bibinfo {author} {\bibfnamefont
  {U.}~\bibnamefont {Parlak}}, \bibinfo {author} {\bibfnamefont
  {J.}~\bibnamefont {Vogel}},\ and\ \bibinfo {author} {\bibfnamefont
  {W.}~\bibnamefont {Kuch}},\ }\bibfield  {title} {\bibinfo {title}
  {{\color{Gray}Steering of magnetic domain walls by single ultrashort laser
  pulses}},\ }\href {https://doi.org/10.1103/PhysRevB.99.214404} {\bibfield
  {journal} {\bibinfo  {journal} {Phys. Rev. B}\ }\textbf {\bibinfo {volume}
  {99}},\ \bibinfo {pages} {214404} (\bibinfo {year}
  {2019}{\natexlab{a}})}\BibitemShut {NoStop}%
\bibitem [{\citenamefont {Jakobs}\ and\ \citenamefont
  {Atxitia}(2022{\natexlab{a}})}]{PhysRevLett.129.037203}%
  \BibitemOpen
  \bibfield  {author} {\bibinfo {author} {\bibfnamefont {F.}~\bibnamefont
  {Jakobs}}\ and\ \bibinfo {author} {\bibfnamefont {U.}~\bibnamefont
  {Atxitia}},\ }\bibfield  {title} {\bibinfo {title} {{\color{Gray}Universal
  Criteria for Single Femtosecond Pulse Ultrafast Magnetization Switching in
  Ferrimagnets}},\ }\href {https://doi.org/10.1103/PhysRevLett.129.037203}
  {\bibfield  {journal} {\bibinfo  {journal} {Phys. Rev. Lett.}\ }\textbf
  {\bibinfo {volume} {129}},\ \bibinfo {pages} {037203} (\bibinfo {year}
  {2022}{\natexlab{a}})}\BibitemShut {NoStop}%
\bibitem [{\citenamefont {Jakobs}\ and\ \citenamefont
  {Atxitia}(2022{\natexlab{b}})}]{PhysRevB.106.134414}%
  \BibitemOpen
  \bibfield  {author} {\bibinfo {author} {\bibfnamefont {F.}~\bibnamefont
  {Jakobs}}\ and\ \bibinfo {author} {\bibfnamefont {U.}~\bibnamefont
  {Atxitia}},\ }\bibfield  {title} {\bibinfo {title} {{\color{Gray}Bridging
  atomistic spin dynamics methods and phenomenological models of single-pulse
  ultrafast switching in ferrimagnets}},\ }\href
  {https://doi.org/10.1103/PhysRevB.106.134414} {\bibfield  {journal} {\bibinfo
   {journal} {Phys. Rev. B}\ }\textbf {\bibinfo {volume} {106}},\ \bibinfo
  {pages} {134414} (\bibinfo {year} {2022}{\natexlab{b}})}\BibitemShut
  {NoStop}%
\bibitem [{\citenamefont {Jakobs}\ and\ \citenamefont
  {Atxitia}(2022{\natexlab{c}})}]{jakobs_exchange-enhancement_2022}%
  \BibitemOpen
  \bibfield  {author} {\bibinfo {author} {\bibfnamefont {F.}~\bibnamefont
  {Jakobs}}\ and\ \bibinfo {author} {\bibfnamefont {U.}~\bibnamefont
  {Atxitia}},\ }\bibfield  {title} {\bibinfo {title}
  {{\color{Gray}Exchange-enhancement of the ultrafast magnetic order dynamics
  in antiferromagnets}},\ }\href {https://arxiv.org/abs/2206.05783} {\bibfield
  {journal} {\bibinfo  {journal} {arXiv:2206.05783}\ } (\bibinfo {year}
  {2022}{\natexlab{c}})}\BibitemShut {NoStop}%
\bibitem [{\citenamefont {Jakobs}\ and\ \citenamefont
  {Atxitia}(2022{\natexlab{d}})}]{jakobs_universal_2022}%
  \BibitemOpen
  \bibfield  {author} {\bibinfo {author} {\bibfnamefont {F.}~\bibnamefont
  {Jakobs}}\ and\ \bibinfo {author} {\bibfnamefont {U.}~\bibnamefont
  {Atxitia}},\ }\bibfield  {title} {\bibinfo {title} {{\color{Gray}Universal
  Criteria for Single Femtosecond Pulse Ultrafast Magnetization Switching in
  Ferrimagnets}},\ }\href {https://doi.org/10.1103/PhysRevLett.129.037203}
  {\bibfield  {journal} {\bibinfo  {journal} {Phys. Rev. Lett.}\ }\textbf
  {\bibinfo {volume} {129}},\ \bibinfo {pages} {037203} (\bibinfo {year}
  {2022}{\natexlab{d}})}\BibitemShut {NoStop}%
\bibitem [{\citenamefont {Jakobs}\ \emph
  {et~al.}(2021{\natexlab{a}})\citenamefont {Jakobs}, \citenamefont {Ostler},
  \citenamefont {Lambert}, \citenamefont {Yang}, \citenamefont {Salahuddin},
  \citenamefont {Wilson}, \citenamefont {Gorchon}, \citenamefont {Bokor},\ and\
  \citenamefont {Atxitia}}]{jakobs_unifying_2021}%
  \BibitemOpen
  \bibfield  {author} {\bibinfo {author} {\bibfnamefont {F.}~\bibnamefont
  {Jakobs}}, \bibinfo {author} {\bibfnamefont {T.~A.}\ \bibnamefont {Ostler}},
  \bibinfo {author} {\bibfnamefont {C.-H.}\ \bibnamefont {Lambert}}, \bibinfo
  {author} {\bibfnamefont {Y.}~\bibnamefont {Yang}}, \bibinfo {author}
  {\bibfnamefont {S.}~\bibnamefont {Salahuddin}}, \bibinfo {author}
  {\bibfnamefont {R.~B.}\ \bibnamefont {Wilson}}, \bibinfo {author}
  {\bibfnamefont {J.}~\bibnamefont {Gorchon}}, \bibinfo {author} {\bibfnamefont
  {J.}~\bibnamefont {Bokor}},\ and\ \bibinfo {author} {\bibfnamefont
  {U.}~\bibnamefont {Atxitia}},\ }\bibfield  {title} {\bibinfo {title}
  {{\color{Gray}Unifying femtosecond and picosecond single-pulse magnetic
  switching in Gd-Fe-Co}},\ }\href
  {https://doi.org/10.1103/PhysRevB.103.104422} {\bibfield  {journal} {\bibinfo
   {journal} {Phys. Rev. B.}\ }\textbf {\bibinfo {volume} {103}},\ \bibinfo
  {pages} {104422} (\bibinfo {year} {2021}{\natexlab{a}})}\BibitemShut
  {NoStop}%
\bibitem [{\citenamefont {Atxitia}\ and\ \citenamefont
  {Ostler}(2018)}]{Atxitia2018}%
  \BibitemOpen
  \bibfield  {author} {\bibinfo {author} {\bibfnamefont {U.}~\bibnamefont
  {Atxitia}}\ and\ \bibinfo {author} {\bibfnamefont {T.~A.}\ \bibnamefont
  {Ostler}},\ }\bibfield  {title} {\bibinfo {title} {{\color{Gray}Ultrafast
  double magnetization switching in GdFeCo with two picosecond-delayed
  femtosecond pump pulses}},\ }\href {https://doi.org/10.1063/1.5044272}
  {\bibfield  {journal} {\bibinfo  {journal} {Appl. Phys. Lett.}\ }\textbf
  {\bibinfo {volume} {113}},\ \bibinfo {pages} {062402} (\bibinfo {year}
  {2018})}\BibitemShut {NoStop}%
\bibitem [{\citenamefont {Radu}\ \emph {et~al.}(2015)\citenamefont {Radu},
  \citenamefont {Stamm}, \citenamefont {Eschenlohr}, \citenamefont {Radu},
  \citenamefont {Abrudan}, \citenamefont {Vahaplar}, \citenamefont {Kachel},
  \citenamefont {Pontius}, \citenamefont {Mitzner}, \citenamefont {Holldack},
  \citenamefont {Föhlisch}, \citenamefont {Ostler}, \citenamefont {Mentink},
  \citenamefont {Evans}, \citenamefont {Chantrell}, \citenamefont {Tsukamoto},
  \citenamefont {Itoh}, \citenamefont {Kirilyuk}, \citenamefont {Kimel},\ and\
  \citenamefont {Rasing}}]{radu_ultrafast_2015}%
  \BibitemOpen
  \bibfield  {author} {\bibinfo {author} {\bibfnamefont {I.}~\bibnamefont
  {Radu}}, \bibinfo {author} {\bibfnamefont {C.}~\bibnamefont {Stamm}},
  \bibinfo {author} {\bibfnamefont {A.}~\bibnamefont {Eschenlohr}}, \bibinfo
  {author} {\bibfnamefont {F.}~\bibnamefont {Radu}}, \bibinfo {author}
  {\bibfnamefont {R.}~\bibnamefont {Abrudan}}, \bibinfo {author} {\bibfnamefont
  {K.}~\bibnamefont {Vahaplar}}, \bibinfo {author} {\bibfnamefont
  {T.}~\bibnamefont {Kachel}}, \bibinfo {author} {\bibfnamefont
  {N.}~\bibnamefont {Pontius}}, \bibinfo {author} {\bibfnamefont
  {R.}~\bibnamefont {Mitzner}}, \bibinfo {author} {\bibfnamefont
  {K.}~\bibnamefont {Holldack}}, \bibinfo {author} {\bibfnamefont
  {A.}~\bibnamefont {Föhlisch}}, \bibinfo {author} {\bibfnamefont {T.~A.}\
  \bibnamefont {Ostler}}, \bibinfo {author} {\bibfnamefont {J.~H.}\
  \bibnamefont {Mentink}}, \bibinfo {author} {\bibfnamefont {R.~F.~L.}\
  \bibnamefont {Evans}}, \bibinfo {author} {\bibfnamefont {R.~W.}\ \bibnamefont
  {Chantrell}}, \bibinfo {author} {\bibfnamefont {A.}~\bibnamefont
  {Tsukamoto}}, \bibinfo {author} {\bibfnamefont {A.}~\bibnamefont {Itoh}},
  \bibinfo {author} {\bibfnamefont {A.}~\bibnamefont {Kirilyuk}}, \bibinfo
  {author} {\bibfnamefont {A.~V.}\ \bibnamefont {Kimel}},\ and\ \bibinfo
  {author} {\bibfnamefont {T.}~\bibnamefont {Rasing}},\ }\bibfield  {title}
  {\bibinfo {title} {{\color{Gray}Ultrafast and Distinct Spin Dynamics in
  Magnetic Alloys}},\ }\href {https://doi.org/10.1142/S2010324715500046}
  {\bibfield  {journal} {\bibinfo  {journal} {SPIN}\ }\textbf {\bibinfo
  {volume} {5}},\ \bibinfo {pages} {1550004} (\bibinfo {year}
  {2015})}\BibitemShut {NoStop}%
\bibitem [{\citenamefont {Liu}\ \emph {et~al.}(2024)\citenamefont {Liu},
  \citenamefont {Weng}, \citenamefont {Song}, \citenamefont {Cai},
  \citenamefont {Tan},\ and\ \citenamefont {Xu}}]{Liu2024}%
  \BibitemOpen
  \bibfield  {author} {\bibinfo {author} {\bibfnamefont {D.}~\bibnamefont
  {Liu}}, \bibinfo {author} {\bibfnamefont {J.}~\bibnamefont {Weng}}, \bibinfo
  {author} {\bibfnamefont {X.}~\bibnamefont {Song}}, \bibinfo {author}
  {\bibfnamefont {W.}~\bibnamefont {Cai}}, \bibinfo {author} {\bibfnamefont
  {S.}~\bibnamefont {Tan}},\ and\ \bibinfo {author} {\bibfnamefont
  {C.}~\bibnamefont {Xu}},\ }\bibfield  {title} {\bibinfo {title}
  {{\color{Gray}Ultrafast write-read event in helicity-independent all-optical
  switching of GdFeCo}},\ }\href {https://doi.org/10.1016/j.jmmm.2024.171824}
  {\bibfield  {journal} {\bibinfo  {journal} {J. Magn. Magn. Mater.}\ }\textbf
  {\bibinfo {volume} {592}},\ \bibinfo {pages} {171824} (\bibinfo {year}
  {2024})}\BibitemShut {NoStop}%
\bibitem [{\citenamefont {Atxitia}\ \emph {et~al.}(2014)\citenamefont
  {Atxitia}, \citenamefont {Barker}, \citenamefont {Chantrell},\ and\
  \citenamefont {Chubykalo-Fesenko}}]{PhysRevB.89.224421}%
  \BibitemOpen
  \bibfield  {author} {\bibinfo {author} {\bibfnamefont {U.}~\bibnamefont
  {Atxitia}}, \bibinfo {author} {\bibfnamefont {J.}~\bibnamefont {Barker}},
  \bibinfo {author} {\bibfnamefont {R.~W.}\ \bibnamefont {Chantrell}},\ and\
  \bibinfo {author} {\bibfnamefont {O.}~\bibnamefont {Chubykalo-Fesenko}},\
  }\bibfield  {title} {\bibinfo {title} {{\color{Gray}Controlling the polarity
  of the transient ferromagneticlike state in ferrimagnets}},\ }\href
  {https://doi.org/10.1103/PhysRevB.89.224421} {\bibfield  {journal} {\bibinfo
  {journal} {Phys. Rev. B}\ }\textbf {\bibinfo {volume} {89}},\ \bibinfo
  {pages} {224421} (\bibinfo {year} {2014})}\BibitemShut {NoStop}%
\bibitem [{\citenamefont {Liu}\ \emph {et~al.}(2023)\citenamefont {Liu},
  \citenamefont {Jiang}, \citenamefont {Wang},\ and\ \citenamefont
  {Xu}}]{liu_minimum_2023}%
  \BibitemOpen
  \bibfield  {author} {\bibinfo {author} {\bibfnamefont {D.}~\bibnamefont
  {Liu}}, \bibinfo {author} {\bibfnamefont {C.}~\bibnamefont {Jiang}}, \bibinfo
  {author} {\bibfnamefont {N.}~\bibnamefont {Wang}},\ and\ \bibinfo {author}
  {\bibfnamefont {C.}~\bibnamefont {Xu}},\ }\bibfield  {title} {\bibinfo
  {title} {{\color{Gray}Minimum separation between two pump pulses for
  ultrafast double magnetization switching in GdFeCo}},\ }\href
  {https://doi.org/10.1063/5.0168770} {\bibfield  {journal} {\bibinfo
  {journal} {Appl. Phys. Lett.}\ }\textbf {\bibinfo {volume} {123}},\ \bibinfo
  {pages} {162401} (\bibinfo {year} {2023})}\BibitemShut {NoStop}%
\bibitem [{\citenamefont {Lawrenz}(2024)}]{Dominic_2024}%
  \BibitemOpen
  \bibfield  {author} {\bibinfo {author} {\bibfnamefont {D.}~\bibnamefont
  {Lawrenz}},\ }\emph {\bibinfo {title} {Ultraschnelle Magnetisierungsdynamik
  ferromagnetischer Schichtsysteme}},\ \href
  {https://doi.org/10.17169/REFUBIUM-45339} {\bibinfo {type} {Phd thesis}},\
  \bibinfo  {school} {Freie Universität Berlin} (\bibinfo {year}
  {2024})\BibitemShut {NoStop}%
\bibitem [{\citenamefont {Ohta}\ and\ \citenamefont
  {Ishida}(1990)}]{ohta_matrix_1990}%
  \BibitemOpen
  \bibfield  {author} {\bibinfo {author} {\bibfnamefont {K.}~\bibnamefont
  {Ohta}}\ and\ \bibinfo {author} {\bibfnamefont {H.}~\bibnamefont {Ishida}},\
  }\bibfield  {title} {\bibinfo {title} {{\color{Gray}Matrix formalism for
  calculation of electric field intensity of light in stratified multilayered
  films}},\ }\href {https://doi.org/10.1364/AO.29.001952} {\bibfield  {journal}
  {\bibinfo  {journal} {Appl. Opt.}\ }\textbf {\bibinfo {volume} {29}},\
  \bibinfo {pages} {1952} (\bibinfo {year} {1990})}\BibitemShut {NoStop}%
\bibitem [{\citenamefont {Chimata}\ \emph {et~al.}(2015)\citenamefont
  {Chimata}, \citenamefont {Isaeva}, \citenamefont {K\'adas}, \citenamefont
  {Bergman}, \citenamefont {Sanyal}, \citenamefont {Mentink}, \citenamefont
  {Katsnelson}, \citenamefont {Rasing}, \citenamefont {Kirilyuk}, \citenamefont
  {Kimel}, \citenamefont {Eriksson},\ and\ \citenamefont
  {Pereiro}}]{PhysRevB.92.094411}%
  \BibitemOpen
  \bibfield  {author} {\bibinfo {author} {\bibfnamefont {R.}~\bibnamefont
  {Chimata}}, \bibinfo {author} {\bibfnamefont {L.}~\bibnamefont {Isaeva}},
  \bibinfo {author} {\bibfnamefont {K.}~\bibnamefont {K\'adas}}, \bibinfo
  {author} {\bibfnamefont {A.}~\bibnamefont {Bergman}}, \bibinfo {author}
  {\bibfnamefont {B.}~\bibnamefont {Sanyal}}, \bibinfo {author} {\bibfnamefont
  {J.~H.}\ \bibnamefont {Mentink}}, \bibinfo {author} {\bibfnamefont {M.~I.}\
  \bibnamefont {Katsnelson}}, \bibinfo {author} {\bibfnamefont
  {T.}~\bibnamefont {Rasing}}, \bibinfo {author} {\bibfnamefont
  {A.}~\bibnamefont {Kirilyuk}}, \bibinfo {author} {\bibfnamefont
  {A.}~\bibnamefont {Kimel}}, \bibinfo {author} {\bibfnamefont
  {O.}~\bibnamefont {Eriksson}},\ and\ \bibinfo {author} {\bibfnamefont
  {M.}~\bibnamefont {Pereiro}},\ }\bibfield  {title} {\bibinfo {title}
  {{\color{Gray}All-thermal switching of amorphous Gd-Fe alloys: Analysis of
  structural properties and magnetization dynamics}},\ }\href
  {https://doi.org/10.1103/PhysRevB.92.094411} {\bibfield  {journal} {\bibinfo
  {journal} {Phys. Rev. B.}\ }\textbf {\bibinfo {volume} {92}},\ \bibinfo
  {pages} {094411} (\bibinfo {year} {2015})}\BibitemShut {NoStop}%
\bibitem [{\citenamefont {Jakobs}\ \emph
  {et~al.}(2021{\natexlab{b}})\citenamefont {Jakobs}, \citenamefont {Ostler},
  \citenamefont {Lambert}, \citenamefont {Yang}, \citenamefont {Salahuddin},
  \citenamefont {Wilson}, \citenamefont {Gorchon}, \citenamefont {Bokor},\ and\
  \citenamefont {Atxitia}}]{PhysRevB.103.104422}%
  \BibitemOpen
  \bibfield  {author} {\bibinfo {author} {\bibfnamefont {F.}~\bibnamefont
  {Jakobs}}, \bibinfo {author} {\bibfnamefont {T.~A.}\ \bibnamefont {Ostler}},
  \bibinfo {author} {\bibfnamefont {C.-H.}\ \bibnamefont {Lambert}}, \bibinfo
  {author} {\bibfnamefont {Y.}~\bibnamefont {Yang}}, \bibinfo {author}
  {\bibfnamefont {S.}~\bibnamefont {Salahuddin}}, \bibinfo {author}
  {\bibfnamefont {R.~B.}\ \bibnamefont {Wilson}}, \bibinfo {author}
  {\bibfnamefont {J.}~\bibnamefont {Gorchon}}, \bibinfo {author} {\bibfnamefont
  {J.}~\bibnamefont {Bokor}},\ and\ \bibinfo {author} {\bibfnamefont
  {U.}~\bibnamefont {Atxitia}},\ }\bibfield  {title} {\bibinfo {title}
  {{\color{Gray}Unifying femtosecond and picosecond single-pulse magnetic
  switching in Gd-Fe-Co}},\ }\href
  {https://doi.org/10.1103/PhysRevB.103.104422} {\bibfield  {journal} {\bibinfo
   {journal} {Phys. Rev. B}\ }\textbf {\bibinfo {volume} {103}},\ \bibinfo
  {pages} {104422} (\bibinfo {year} {2021}{\natexlab{b}})}\BibitemShut
  {NoStop}%
\bibitem [{\citenamefont {Barker}\ \emph {et~al.}(2013)\citenamefont {Barker},
  \citenamefont {Atxitia}, \citenamefont {Ostler}, \citenamefont {Hovorka},
  \citenamefont {Chubykalo-Fesenko},\ and\ \citenamefont
  {Chantrell}}]{barker_two-magnon_2013}%
  \BibitemOpen
  \bibfield  {author} {\bibinfo {author} {\bibfnamefont {J.}~\bibnamefont
  {Barker}}, \bibinfo {author} {\bibfnamefont {U.}~\bibnamefont {Atxitia}},
  \bibinfo {author} {\bibfnamefont {T.~A.}\ \bibnamefont {Ostler}}, \bibinfo
  {author} {\bibfnamefont {O.}~\bibnamefont {Hovorka}}, \bibinfo {author}
  {\bibfnamefont {O.}~\bibnamefont {Chubykalo-Fesenko}},\ and\ \bibinfo
  {author} {\bibfnamefont {R.~W.}\ \bibnamefont {Chantrell}},\ }\bibfield
  {title} {\bibinfo {title} {{\color{Gray}Two-magnon bound state causes
  ultrafast thermally induced magnetisation switching}},\ }\href
  {https://doi.org/10.1038/srep03262} {\bibfield  {journal} {\bibinfo
  {journal} {Sci Rep.}\ }\textbf {\bibinfo {volume} {3}},\ \bibinfo {pages}
  {3262} (\bibinfo {year} {2013})}\BibitemShut {NoStop}%
\bibitem [{\citenamefont {Shokr}\ \emph
  {et~al.}(2019{\natexlab{b}})\citenamefont {Shokr}, \citenamefont {Sandig},
  \citenamefont {Erkovan}, \citenamefont {Zhang}, \citenamefont {Bernien},
  \citenamefont {{\"U}nal}, \citenamefont {Kronast}, \citenamefont {Parlak},
  \citenamefont {Vogel},\ and\ \citenamefont {Kuch}}]{shokr_steering_2019}%
  \BibitemOpen
  \bibfield  {author} {\bibinfo {author} {\bibfnamefont {Y.~A.}\ \bibnamefont
  {Shokr}}, \bibinfo {author} {\bibfnamefont {O.}~\bibnamefont {Sandig}},
  \bibinfo {author} {\bibfnamefont {M.}~\bibnamefont {Erkovan}}, \bibinfo
  {author} {\bibfnamefont {B.}~\bibnamefont {Zhang}}, \bibinfo {author}
  {\bibfnamefont {M.}~\bibnamefont {Bernien}}, \bibinfo {author} {\bibfnamefont
  {A.~A.}\ \bibnamefont {{\"U}nal}}, \bibinfo {author} {\bibfnamefont
  {F.}~\bibnamefont {Kronast}}, \bibinfo {author} {\bibfnamefont
  {U.}~\bibnamefont {Parlak}}, \bibinfo {author} {\bibfnamefont
  {J.}~\bibnamefont {Vogel}},\ and\ \bibinfo {author} {\bibfnamefont
  {W.}~\bibnamefont {Kuch}},\ }\bibfield  {title} {\bibinfo {title}
  {{\color{Gray}Steering of magnetic domain walls by single ultrashort laser
  pulses}},\ }\href {https://doi.org/10.1103/PhysRevB.99.214404} {\bibfield
  {journal} {\bibinfo  {journal} {Phys. Rev. B}\ }\textbf {\bibinfo {volume}
  {99}},\ \bibinfo {pages} {214404} (\bibinfo {year}
  {2019}{\natexlab{b}})}\BibitemShut {NoStop}%
\bibitem [{\citenamefont {Steinbach}\ \emph {et~al.}(2024)\citenamefont
  {Steinbach}, \citenamefont {Atxitia}, \citenamefont {Yao}, \citenamefont
  {Borchert}, \citenamefont {Engel}, \citenamefont {Bencivenga}, \citenamefont
  {Foglia}, \citenamefont {Mincigrucci}, \citenamefont {Pedersoli},
  \citenamefont {De~Angelis}, \citenamefont {Pancaldi}, \citenamefont
  {Fainozzi}, \citenamefont {Pelli~Cresi}, \citenamefont {Paltanin},
  \citenamefont {Capotondi}, \citenamefont {Masciovecchio}, \citenamefont
  {Eisebitt},\ and\ \citenamefont {von
  Korff~Schmising}}]{steinbach_exploring_2024}%
  \BibitemOpen
  \bibfield  {author} {\bibinfo {author} {\bibfnamefont {F.}~\bibnamefont
  {Steinbach}}, \bibinfo {author} {\bibfnamefont {U.}~\bibnamefont {Atxitia}},
  \bibinfo {author} {\bibfnamefont {K.}~\bibnamefont {Yao}}, \bibinfo {author}
  {\bibfnamefont {M.}~\bibnamefont {Borchert}}, \bibinfo {author}
  {\bibfnamefont {D.}~\bibnamefont {Engel}}, \bibinfo {author} {\bibfnamefont
  {F.}~\bibnamefont {Bencivenga}}, \bibinfo {author} {\bibfnamefont
  {L.}~\bibnamefont {Foglia}}, \bibinfo {author} {\bibfnamefont
  {R.}~\bibnamefont {Mincigrucci}}, \bibinfo {author} {\bibfnamefont
  {E.}~\bibnamefont {Pedersoli}}, \bibinfo {author} {\bibfnamefont
  {D.}~\bibnamefont {De~Angelis}}, \bibinfo {author} {\bibfnamefont
  {M.}~\bibnamefont {Pancaldi}}, \bibinfo {author} {\bibfnamefont
  {D.}~\bibnamefont {Fainozzi}}, \bibinfo {author} {\bibfnamefont {J.~S.}\
  \bibnamefont {Pelli~Cresi}}, \bibinfo {author} {\bibfnamefont
  {E.}~\bibnamefont {Paltanin}}, \bibinfo {author} {\bibfnamefont
  {F.}~\bibnamefont {Capotondi}}, \bibinfo {author} {\bibfnamefont
  {C.}~\bibnamefont {Masciovecchio}}, \bibinfo {author} {\bibfnamefont
  {S.}~\bibnamefont {Eisebitt}},\ and\ \bibinfo {author} {\bibfnamefont
  {C.}~\bibnamefont {von Korff~Schmising}},\ }\bibfield  {title} {\bibinfo
  {title} {{\color{Gray}Exploring the Fundamental Spatial Limits of Magnetic
  All-Optical Switching}},\ }\href
  {https://doi.org/10.1021/acs.nanolett.4c00129} {\bibfield  {journal}
  {\bibinfo  {journal} {Nano Lett.}\ }\textbf {\bibinfo {volume} {24}},\
  \bibinfo {pages} {6865} (\bibinfo {year} {2024})}\BibitemShut {NoStop}%
\end{thebibliography}
%

\end{document}


\title{Supplemental Material for Studying all-optical magnetization switching of GdFe by double-pulse laser excitation}
\author{\surname{Rahil} Hosseinifar\orcidlink{0000-0002-9124-008X}}%
\affiliation{Institut f\"ur Experimentalphysik, Freie Universit\"at Berlin, Arnimallee 14, 14195 Berlin, Germany}
\author{\surname{Felix} Steinbach\orcidlink{0000-0001-9287-1062}}%
\affiliation{Max-Born-Institut f\"ur Nichtlineare Optik und Kurzzeitspektroskopie, Max-Born-Stra{\ss}e 2A, 12489 Berlin, Germany}
\author{\surname{Ivar} Kumberg\orcidlink{0000-0002-3914-0604}}%
\affiliation{Institut f\"ur Experimentalphysik, Freie Universit\"at Berlin, Arnimallee 14, 14195 Berlin, Germany}
\author{\surname{José}  Miguel Lendínez\orcidlink{}}%
\affiliation{Instituto de Ciencia de Materiales de Madrid, CSIC, Cantoblanco, 28049 Madrid, Spain}
\author{\surname{Sangeeta} Thakur\orcidlink{0000-0003-4879-5650}}%
\author{\surname{Sebastien E.} Hadjadj\orcidlink{0000-0002-6045-574X}}%
\author{\surname{Jendrik } Gördes\orcidlink{}}%
\author{\surname{Chowdhury}  S. Awsaf\orcidlink{0009-0007-4709-6168}}%
\affiliation{Institut f\"ur Experimentalphysik, Freie Universit\"at Berlin, Arnimallee 14, 14195 Berlin, Germany}
\author{\surname{Mario} Fix\orcidlink{0000-0002-6677-2674}}%
\affiliation{Institute of Physics, University of Augsburg, Universit\"atsstra{\ss}e 1, 86135 Augsburg, Germany}
\author{\surname{Manfred} Albrecht\orcidlink{0000-0002-0795-8487}}%
\affiliation{Institute of Physics, University of Augsburg, Universit\"atsstra{\ss}e 1, 86135 Augsburg, Germany}
\author{\surname{Florian} Kronast\orcidlink{0000-0001-6048-480X}}%
\affiliation{Helmholtz-Zentrum Berlin f\"ur Materialien und Energie, Albert-Einstein-Stra{\ss}e 15, 12489 Berlin, Germany}
\author{\surname{Unai} Atxitia\orcidlink{0000-0002-2871-5644}}%
\affiliation{Instituto de Ciencia de Materiales de Madrid, CSIC, Cantoblanco, 28049 Madrid, Spain}
\author{\surname{Clemens} von Korff Schmising\orcidlink{0000-0003-3159-3489}}%
\affiliation{Max-Born-Institut f\"ur Nichtlineare Optik und Kurzzeitspektroskopie, Max-Born-Stra{\ss}e 2A, 12489 Berlin, Germany}
\author{\surname{Wolfgang} Kuch\orcidlink{0000-0002-5764-4574}}%
\affiliation{Institut f\"ur Experimentalphysik, Freie Universit\"at Berlin, Arnimallee 14, 14195 Berlin, Germany}
\email{Correspondence and requests for materials should be addressed to W.K. (email: kuch@physik.fu-berlin.de)}
\affiliation{Institut f\"ur Experimentalphysik, Freie Universit\"at Berlin, Arnimallee 14, 14195 Berlin, Germany}
\date{\today}
\maketitle

\section*{Absorbed fluence of the laser}
 In the main text,  all the numbers of fluences refer to the the absorbed fluence in the GdFe layer. The fluence absorption of the laser pulse in each individual layer of the sample is calculated using Abeles' formalism \cite{ohta_matrix_1990}. Abele demonstrated the connection between the intensity of the incident electric field wave, the reflected wave $(E\textsubscript{0}^{+}$), and the transmitted wave $(E\textsubscript{0}^{-})$ after passing through $m$ layers (E\textsubscript{m+1}\textsuperscript{+}).  The details of the calculation can be found in Ref.\ \cite{ohta_matrix_1990}.
According to  Poynting's theorem \cite{ohta_matrix_1990}, the absorption at a certain depth $z$ in the sample is proportional to the intensity of the electric field of the light at that depth, such that the absorption in the thickness range between specific depths $z\textsubscript{1}<z\textsubscript{2}$ can be calculated as:

 \begin{equation}
 \begin{split}
 dA = q_j F(z) dz\\
A(z_1< z < z_2) = \int_{z_1}^{z_2} q_j F(z) dz  
 \end{split}
 \end{equation}
 
 $A$ is the absorption, $F(z)$ stands for the electric field intensity at depth $z$, $j$ labels the different layers, and $q$ is defined as \cite{ohta_matrix_1990}:
 
 \begin{equation}
 \begin{split}
  q_j= Re \left(\frac{n_{j}\cos\theta_{j}}{n_{0}\cos\theta_{0}}\right)\beta_{j}\\
\beta_{j}=2Im(K_{jz})\\
= 2\pi\nu Im(n_{j}\cos\theta_{j})
 \end{split}
 \end{equation}

In this equation, $K_{zj}$ represents the $z$ component of the wave vector, $\theta_0$ is the incident angle, $\theta_j$ the propagation angle in layer $j$, and $n$ represents the index of refraction of layer $j$.
Using this formula, the absorption per layer is calculated.  Table \ref{tab1} presents the indices of refraction that are used in these calculations as well as the resulting absorption in each layer.

\begin{table}[h!]
\centering
\begin{tabular}{ |c|c|c|c|c| } 
 \hline
Layer & thickness (\SI{}{\nano\meter}) & Refractive index & absorption (\%) for $\theta$ =  \SI{76}{\degree}  & absorption (\%) for $\theta$ = \SI{16}{\degree} \\ 
\hline
Vacuum & -- & 1 & -- &--\\ 
Al & 3 & 2.76 + i 8.35 & 10 & 20  \\ 
GdFe & 10 & 2.66 + i 3.6 & 13 & 23\\ 
Pt & 5 & 2.98 + i 6.37 &7 & 14\\ 
SiO & 100 & 1.93 + i 0.00005 & $7 \times 10^{-4}$ & $1.5 \times 10^{-3}$ \\ 
Si & $5 \times 10^{6}$ & 3.696 + i 0.0047023 & 7 & 16\\ 
Vacuum & -- & 1 & -- & --\\ 
 \hline
 \end{tabular}
\caption{
\label{tab1}
The refractive indices used for the different layers of the sample.}
\end{table}
 
\begin{figure}[htb!]
    \centering
    \includegraphics[width=0.8\textwidth]{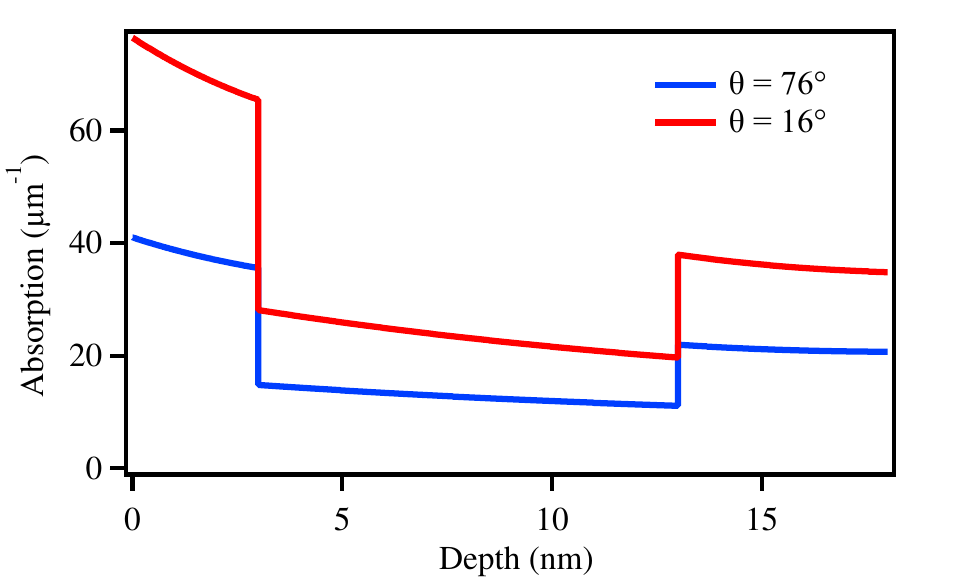} 
    \caption{Optical absorption at \SI{800}{\nano\meter} along the thickness of the sample. The red (blue) lines show the absorption of the sample for the angle of \SI{16}{\degree} (\SI{76}{\degree}) incidence with respect to the sample surface, respectively.}
    \label{figure-s1}
\end{figure}

Figure \ref{figure-s1} shows the absorption of the light as a function of depth across the sample.  The light absorbed in the GdFe layer is reported in the paper as absorbed fluence.  As mentioned before, the angles at which the laser light arrives at the sample are different in the two optical setups.  Both angles are shown in the graph.

\subsection*{Threshold of single-pulse switching}
\begin{figure}[t!]
    \centering
    \includegraphics[width=0.8\textwidth]{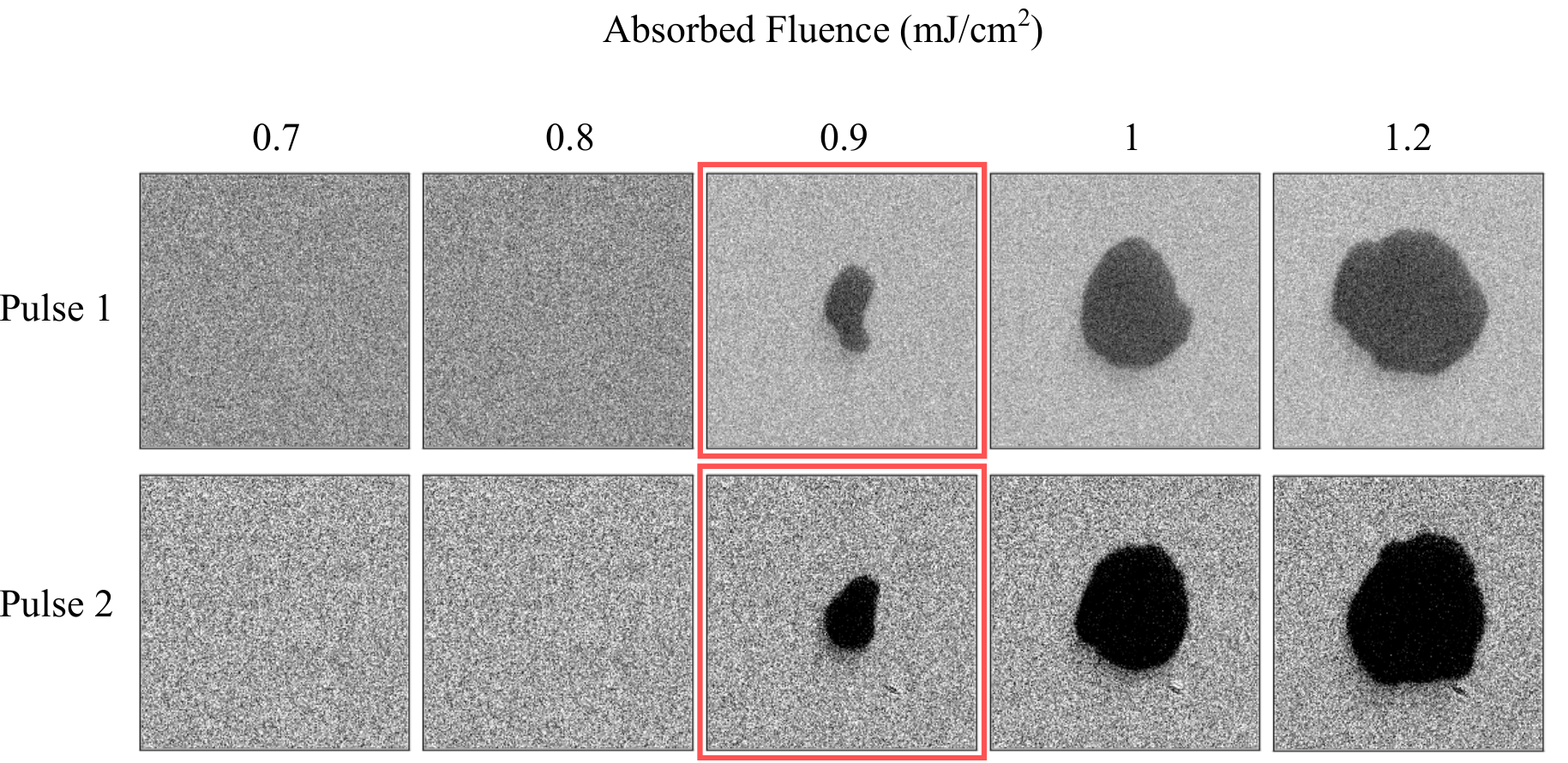}
    \caption{Kerr microscopy images after single-pulse excitation, where always one of the pulses has been blocked. On top of each image the absorbed fluence of the laser pulses is written. Dark and light gray contrast correspond to opposite directions of magnetization. The switching of the magnetization starts at \SI{0.91}{\milli\joule\per\centi\meter\textsuperscript{2}} for both the first and the second pulse. This fluence is assigned as threshold of single-pulse switching. The field of view is $50 \times \SI{50}{\micro\meter\textsuperscript{2}}$.}
    \label{figure-S2}
\end{figure}

\begin{figure}[tb!]
    \centering
    \includegraphics[width=0.8\textwidth]{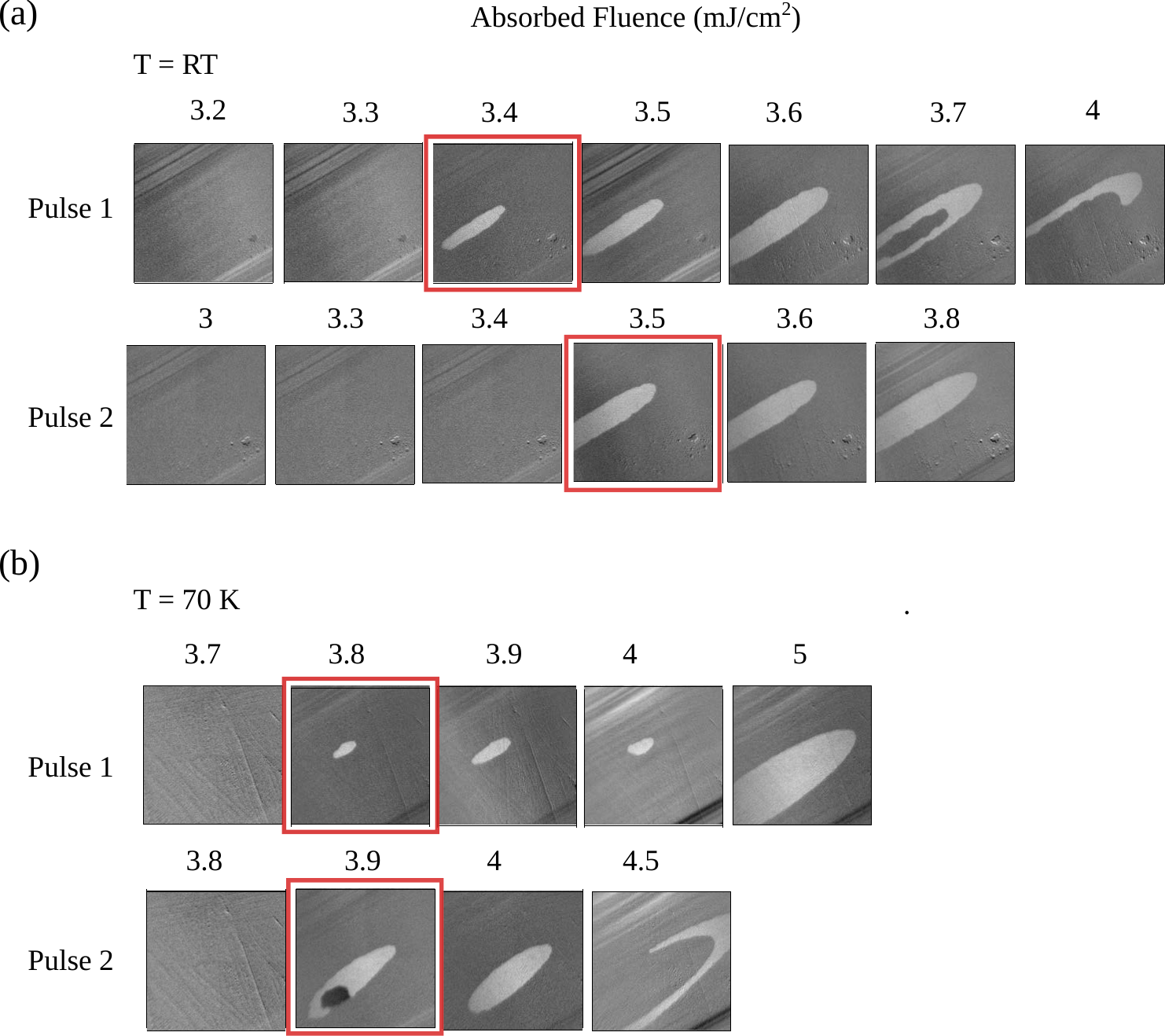}
    \caption{PEEM images for single-pulse switching, where always one of the pulses has been blocked, at (a) room temperature and (b) $T = \SI{70}{\kelvin}$. The absorbed fluence of the laser in the GdFe layer is written on top of each image. The red frames identify the fluences that are defined as threshold of single-pulse switching. The field of view is $20 \times \SI{20}{\micro\meter\textsuperscript{2}}$. }
    \label{figure-s3}
\end{figure}

\noindent The single-pulse switching observed at the Kerr microscopy setup is shown in Fig.\ \ref{figure-S2}.
The threshold of the single-pulse switching for each pulse is defined as the minimum absorbed fluence at which the magnetization switches and a change in the contrast is observed. Both laser pulses start to imply changes in the sample when their fluence reaches \SI{0.9}{\milli\joule\per\centi\meter\textsuperscript{2}}, which is the threshold fluence (F\_{t1}).
The single-pulse switching at the PEEM setup is shown in Fig.\ \ref{figure-s3} for two different temperatures. As can be seen, the threshold for single-pulse switching is higher at \SI{70}{\kelvin} than at room temperature.

\section*{Double-pulse excitation of the sample with different fluence}
\begin{figure}[h!]
    \centering
     \includegraphics[width=0.9\textwidth]{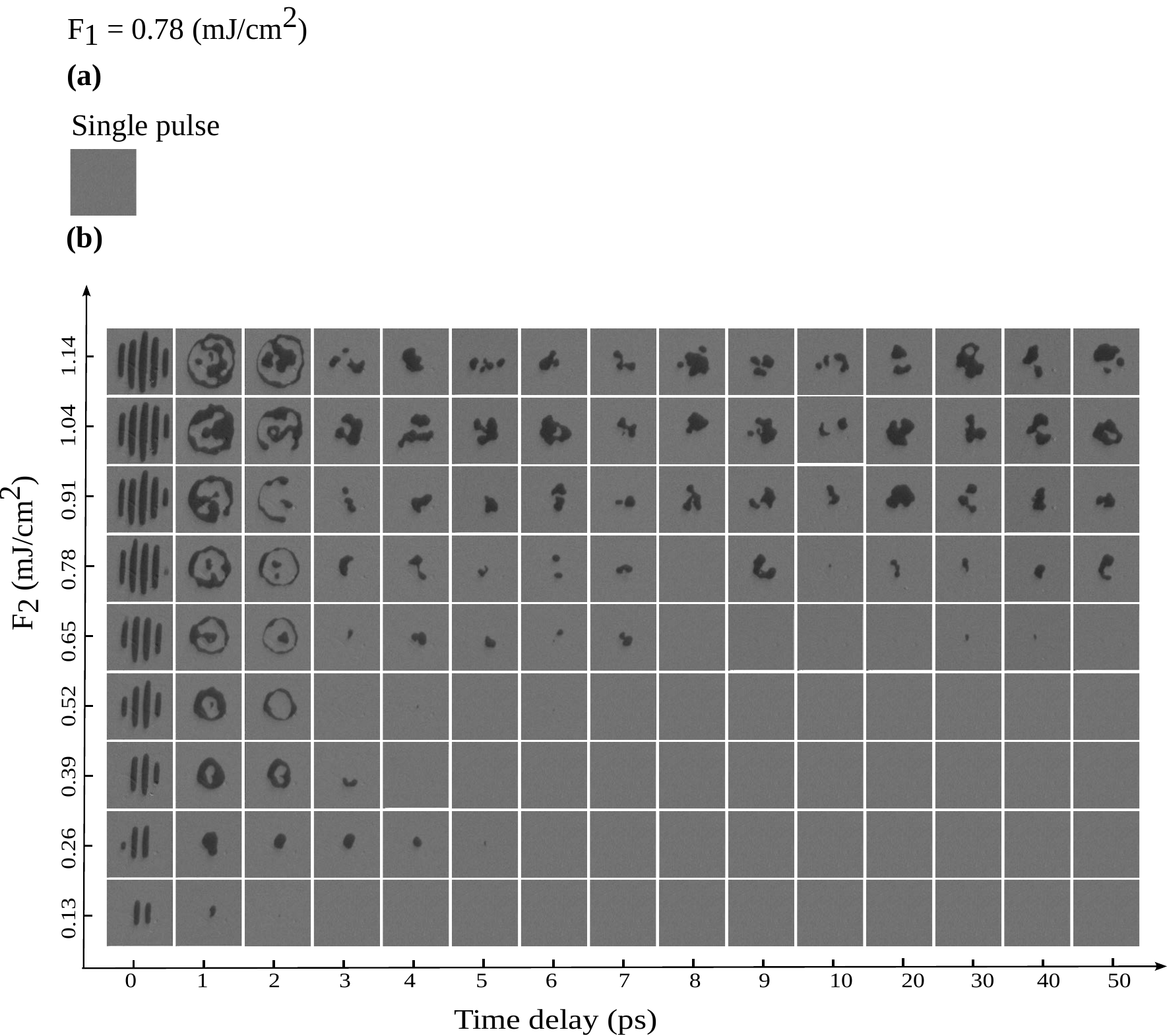}
    \caption{Kerr microscopy images after double-pulse excitation for $F_{1} = \SI{0.78}{\milli\joule\per\centi\meter\textsuperscript{2}}$ at room temperature. Field of view is $50 \times \SI{50}{\micro\meter\textsuperscript{2}}$.}
    \label{figure-S4}
\end{figure}

\begin{figure}[h!]
    \centering
    \includegraphics[width=0.9\textwidth]{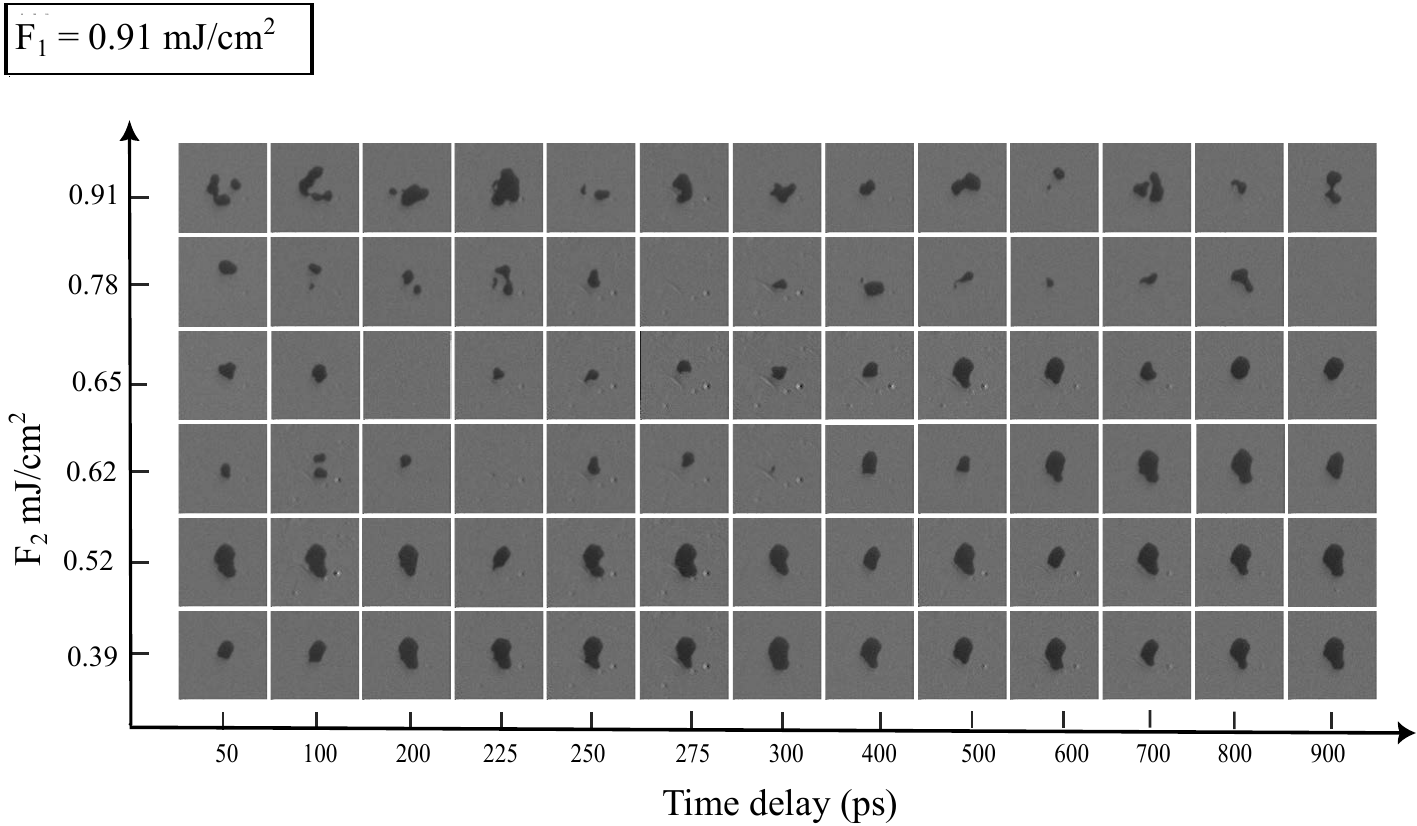}
    \caption{Kerr microscopy images after double-pulse excitation for $F_{1} = \SI{0.91}{\milli\joule\per\centi\meter\textsuperscript{2}}$ at room temperature. Field of view is $50 \times \SI{50}{\micro\meter\textsuperscript{2}}$.}
    \label{figure-S5}
\end{figure}

\begin{figure}[h!]
    \centering
    \includegraphics[width=0.9\textwidth]{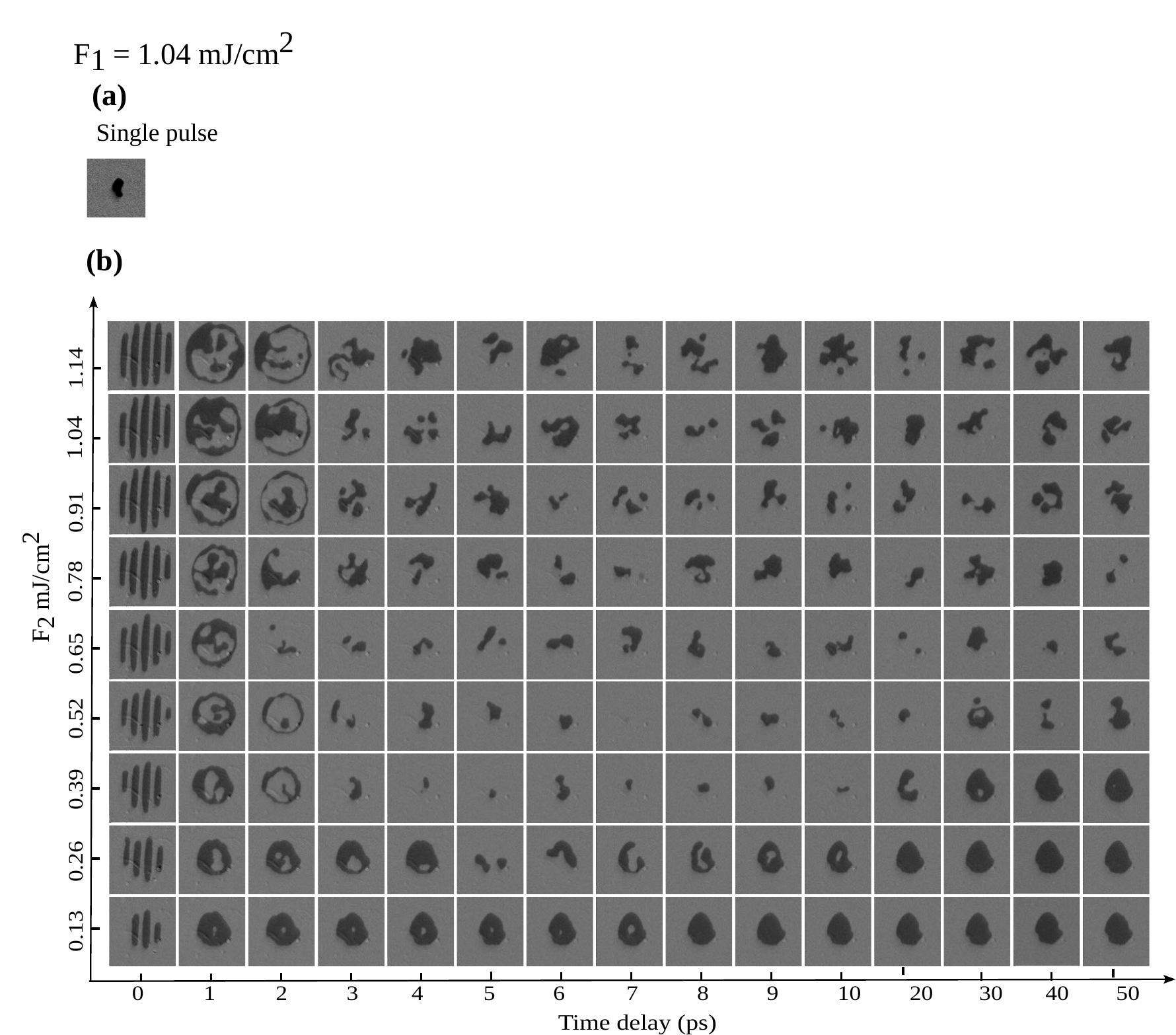}
    \caption{Kerr microscopy images after double-pulse excitation for $F_{1} = \SI{1.04}{\milli\joule\per\centi\meter\textsuperscript{2}}$ at room temperature. Field of view is $50 \times \SI{50}{\micro\meter\textsuperscript{2}}$.}
    \label{figure-S6}
\end{figure}

The magnetization state of the sample after double-pulse excitation for three different fluences $F_1$ of the first pulse is shown in Fig.\  2 in the main text.  Here, each of these fluences is presented individually, with higher fluences of the second pulse to observe the sample's behavior for increased $F_2$. In Fig.\ \ref{figure-S4}, the final state of magnetization for $F_{1} = \SI{0.78}{\milli\joule\per\centi\meter\textsuperscript{2}}$ is displayed.  When $F_2$ exceeds the switching threshold, the sample demagnetizes and forms multi-domains.
This effect is observed for $F_{1} = \SI{0.91}{\milli\joule\per\centi\meter\textsuperscript{2}}$ and $F_{1} = \SI{1.04}{\milli\joule\per\centi\meter\textsuperscript{2}}$ in Figs.\ \ref{figure-S5} and \ref{figure-S6}, respectively.

\subsection*{Double-pulse switching at constant time delays}
\begin{figure}[htb!]
    \centering
    \includegraphics[width=0.8\textwidth]{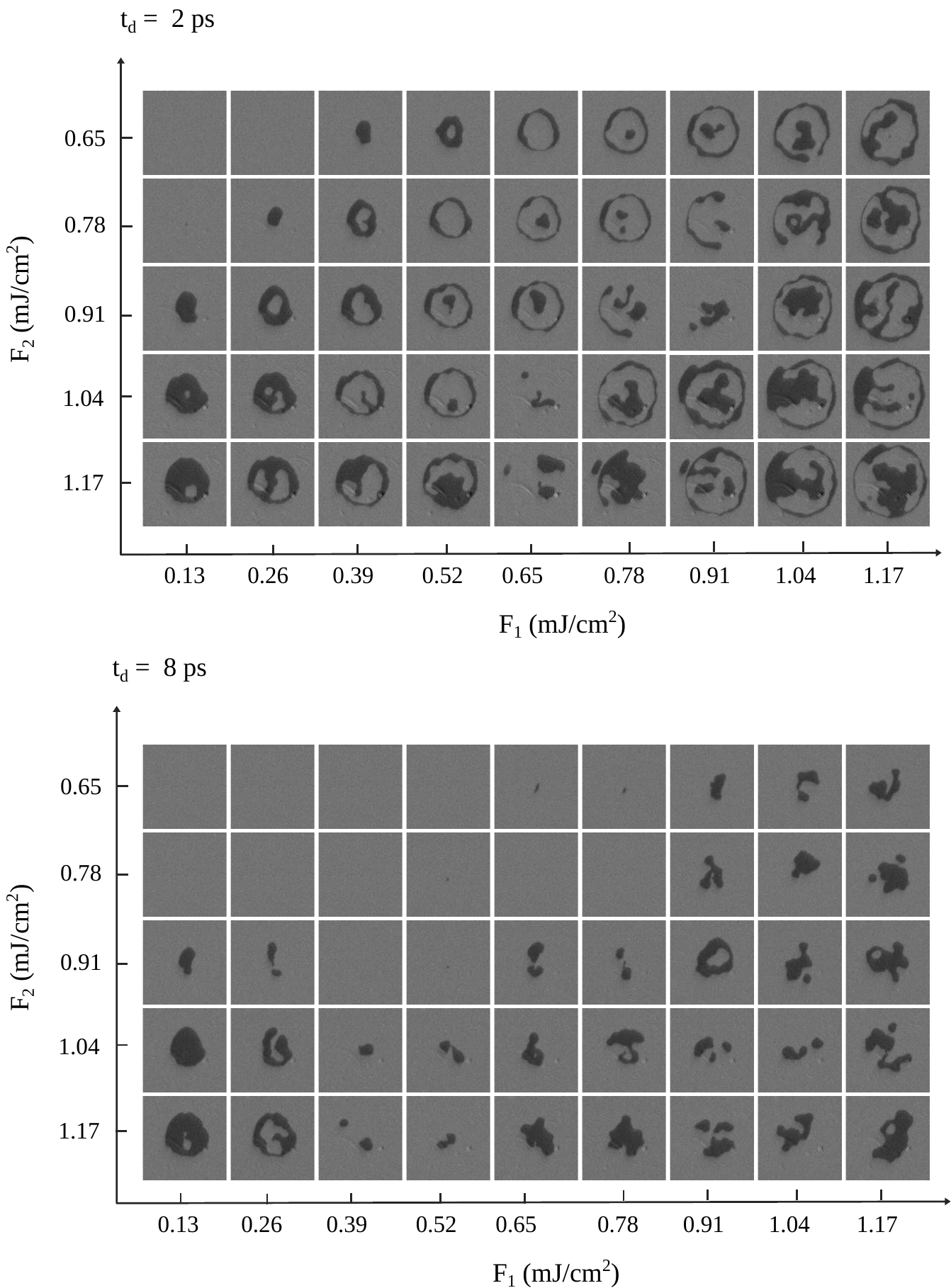}
    \caption{Kerr microscopy images of GdFe after double-pulse excitation for (a) $t_{d} = \SI{2}{\pico\second}$, (b) $t_{d} = \SI{8}{\pico\second}$ at room temperature. Field of view is $50 \times \SI{50}{\micro\meter\textsuperscript{2}}$.}
    \label{figure-S7}
\end{figure}

Fig.\ \ref{figure-S7} shows Kerr microscopy images after double-pulse excitation at $t_{d} = \SI{2}{\pico\second}$, where the sample does not re-switch, and at $t_{d} = \SI{8}{\pico\second}$, where the sample shows re-switching at specific fluences.

\begin{figure}[htb!]
    \centering
    \includegraphics[width=0.9\textwidth]{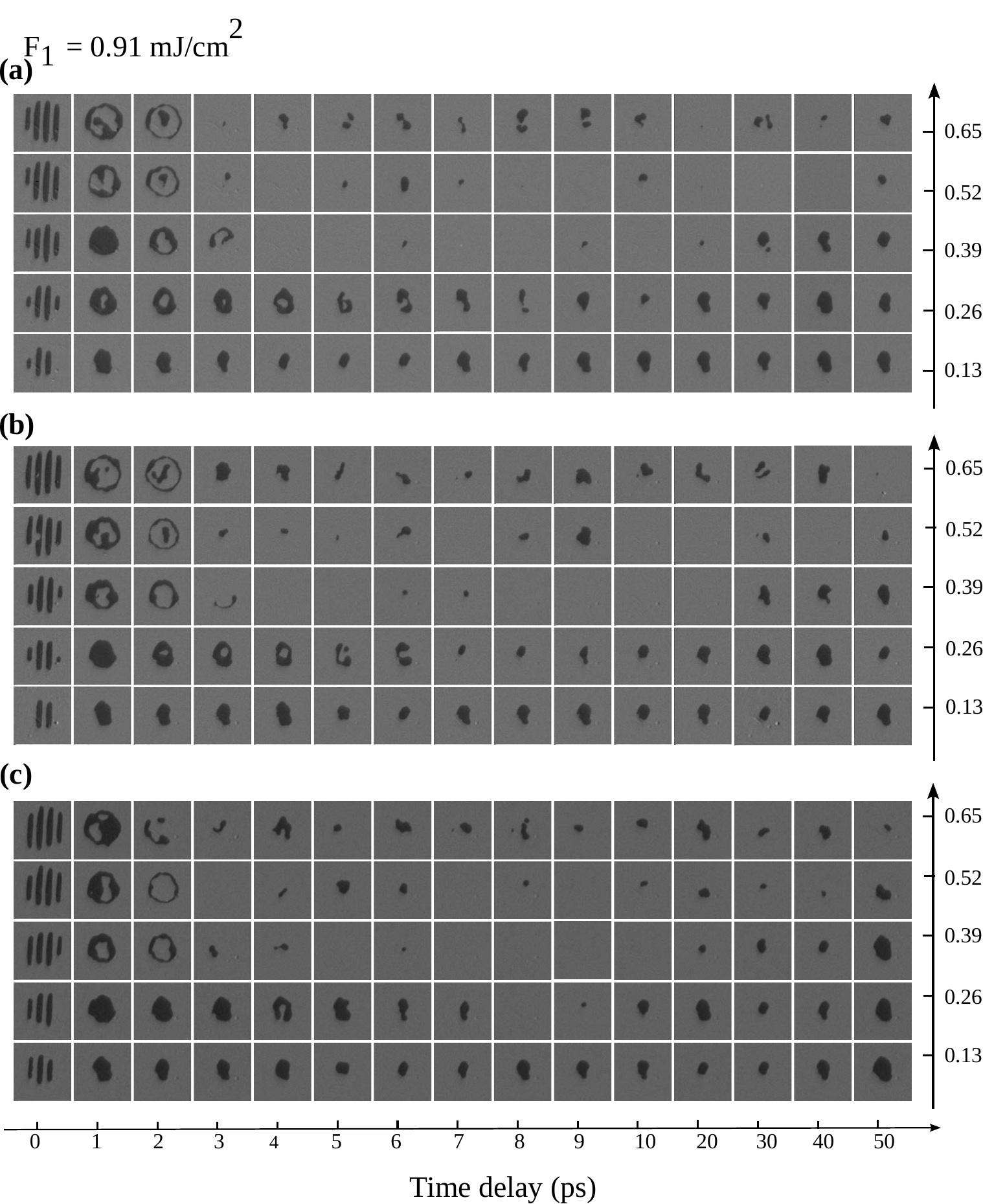}
   \caption{Kerr microscopy images at room temperature for F\textsubscript{1} = \SI{0.91}{\milli\joule\per\centi\meter\textsuperscript{2}}.
   (a-c) The horizontal axis shows time delays, and the left axis shows the fluence of the second pulse.  Field of view is 50$\times$50 \SI{}{\micro\meter\textsuperscript{2}}.}
    \label{secondtime}
\end{figure}

We repeated the experiment multiple times.  The switching behavior of the sample after three repetitions of the experiment is displayed in Fig. \ref{secondtime}.  As shown, when  F\textsubscript{2} = \SI{0.39}{\milli\joule\per\centi\meter\textsuperscript{2}}, the three panels
 indicate that starting at \SI{4}{\pico\second}, the contrast of the reversed magnetization diminishes, indicating re-switching.  This re-switching persists until a time delay of \SI{50}{\pico\second}, after which the sample transitions to toggle-switching for longer delays. 
  Increasing F\textsubscript{2} to \SI{0.52}{\milli\joule\per\centi\meter\textsuperscript{2}}, expands the time-delay window where re-switching is observed in all three panels, and re-switching continuing until a delay of \SI{40}{\pico\second}. The top row of each
   panel presents the results for F\textsubscript{2} = \SI{0.65}{\milli\joule\per\centi\meter\textsuperscript{2}}, where multidomain formation occurs. Re-switching is no longer possible at this fluence level, and the sample undergoes demagnetization instead.

\begin{figure}[htb!]
    \centering
    \includegraphics[width=0.9\textwidth]{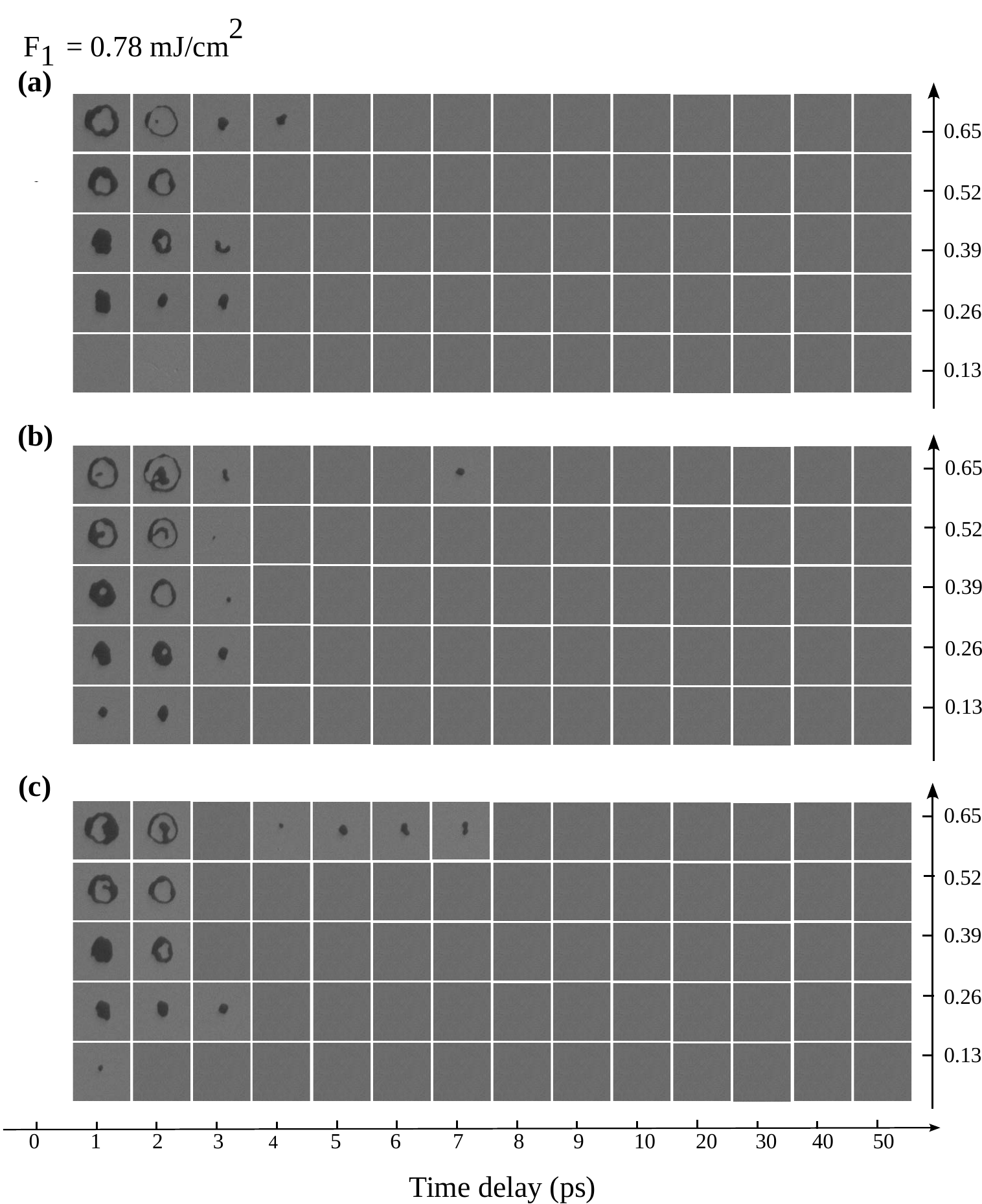}
   \caption{Kerr microscopy images at room temperature for F\textsubscript{1} = \SI{0.78}{\milli\joule\per\centi\meter\textsuperscript{2}}.
   (a-c) The horizontal axis shows time delays, and the left axis shows the fluence of the second pulse.  Field of view is 50$\times$50 \SI{}{\micro\meter\textsuperscript{2}}.}
    \label{Figure-S8R}
\end{figure}

The switching behavior of the sample after three repetitions of the experiment for F\textsubscript{1} = \SI{0.78}{\milli\joule\per\centi\meter\textsuperscript{2}}, where the first pulse is below the switching threshold, is shown in Fig. \ref{Figure-S8R}.
In the top row of each panel, where  F\textsubscript{2} = \SI{0.65}{\milli\joule\per\centi\meter\textsuperscript{2}}, switching persists for  longer time delays.

\section*{Simulation of magnetization dynamics}
\begin{figure}[htb!]
    \centering
    \includegraphics[width=0.9\textwidth]{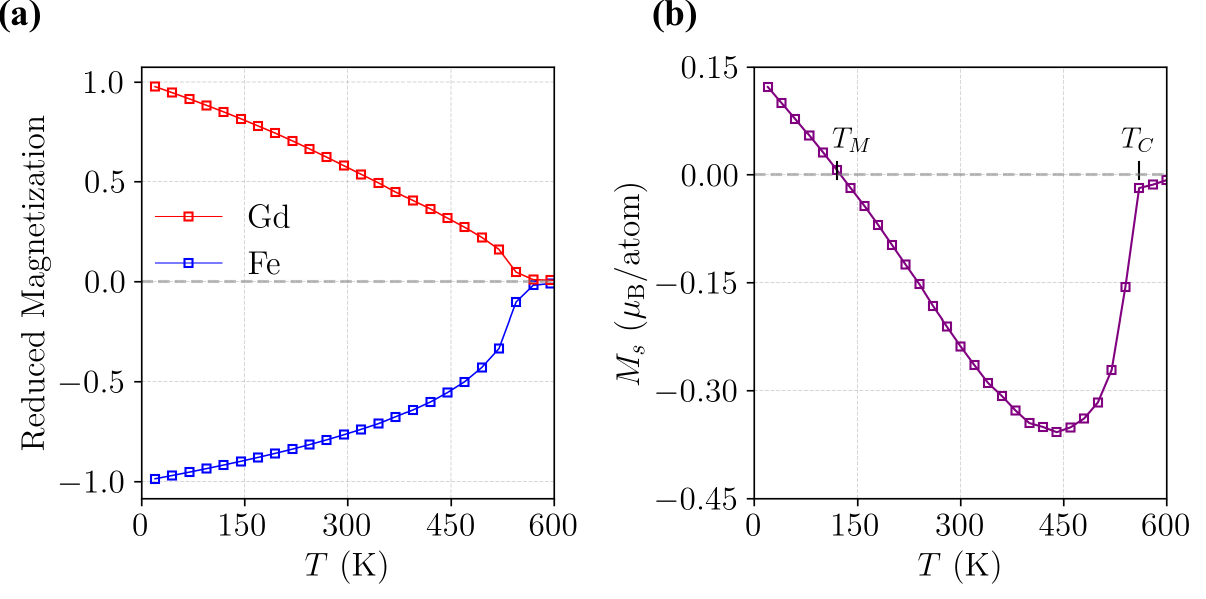}
    \caption{ (a) Calculated magnetization of the Fe and Gd sublattices as a function of temperature. (b) Total calculated magnetization curve of GdFe. Symbols present the computational results, while the dotted lines are used as a visual guide.} 
    \label{figure-S8}
\end{figure}

In the ASD simulations, the atomic species of each spin of the Gd\textsubscript{26}Fe\textsubscript{74} alloy
are randomly located in the lattice structure. The dynamics of each atomic spin follows the
stochastic Landau-Lifshitz-Gilbert (LLG) equation:

\begin{equation}
  \frac{\partial{\textbf s_i}}{\partial{t}} = \frac{|\gamma|}{(1+\alpha_{i}^{2})\mu_{i}}\left[( \textbf s_{i} \times\textbf  H_{i}) - \alpha_{i}(\textbf s_{i}\times(\textbf s_{i}\times \textbf  H_{i}))\right]
\end{equation}

where $\alpha_{i}$ is the local intrinsic atomic damping. The effective field is $ \textbf H_{i}$ = $\boldsymbol{\zeta}_{i}$-$\frac{\partial \textbf{H}}{\partial \textbf{s}_{i}}$, 
where thermal fluctuations are represented by the stochastic field $\boldsymbol{\zeta}_{i}$.  We set the atomic magnetic moments to $\mu\textsubscript{Fe} = \SI{2.4}{} \mu_{B}$
and $\mu\textsubscript{Gd} = \SI{7.38}{} \mu_{B}$, where $\mu_{B}$ is the Bohr magneton as in \cite{PhysRevB.92.094411} and the exchange coupling parameters to
 $J\textsubscript{Fe--Fe} = \SI{3.460}{} \times 10^{-21} \SI{}{\joule}$, $J\textsubscript{Gd--Gd} = \SI{1.350}{} \times 10^{-21} \SI{}{\joule}$, and $J_{Fe-Gd}= \SI{-1.205}{} \times 10^{-21} \SI{}{\joule}$.
 These values result in a compensation temperature next to $T_{M}= \SI{120}{\kelvin}$, mirroring that of the experiment, and a Curie temperature around $T_{C} = \SI{540}{\kelvin}$, as Fig.\ \ref{figure-S8} shows. Additionally, intrinsic damping parameters $\alpha \textsubscript{Fe} = \SI{0.06}{}$, $\alpha \textsubscript{Gd} = \SI{0.01}{}$ are taken from Ref.\ \cite{PhysRevB.103.104422}.
 The on-site uniaxial anisotropy is defined as $d_{i}^{z}$ , with a value of $d_{i}^{z} = \SI{8.010}{}\times 10^{-24}$ \SI{}{\joule} \cite{barker_two-magnon_2013}.

\begin{figure}[htb!]
    \centering
    \includegraphics[width=0.9\textwidth]{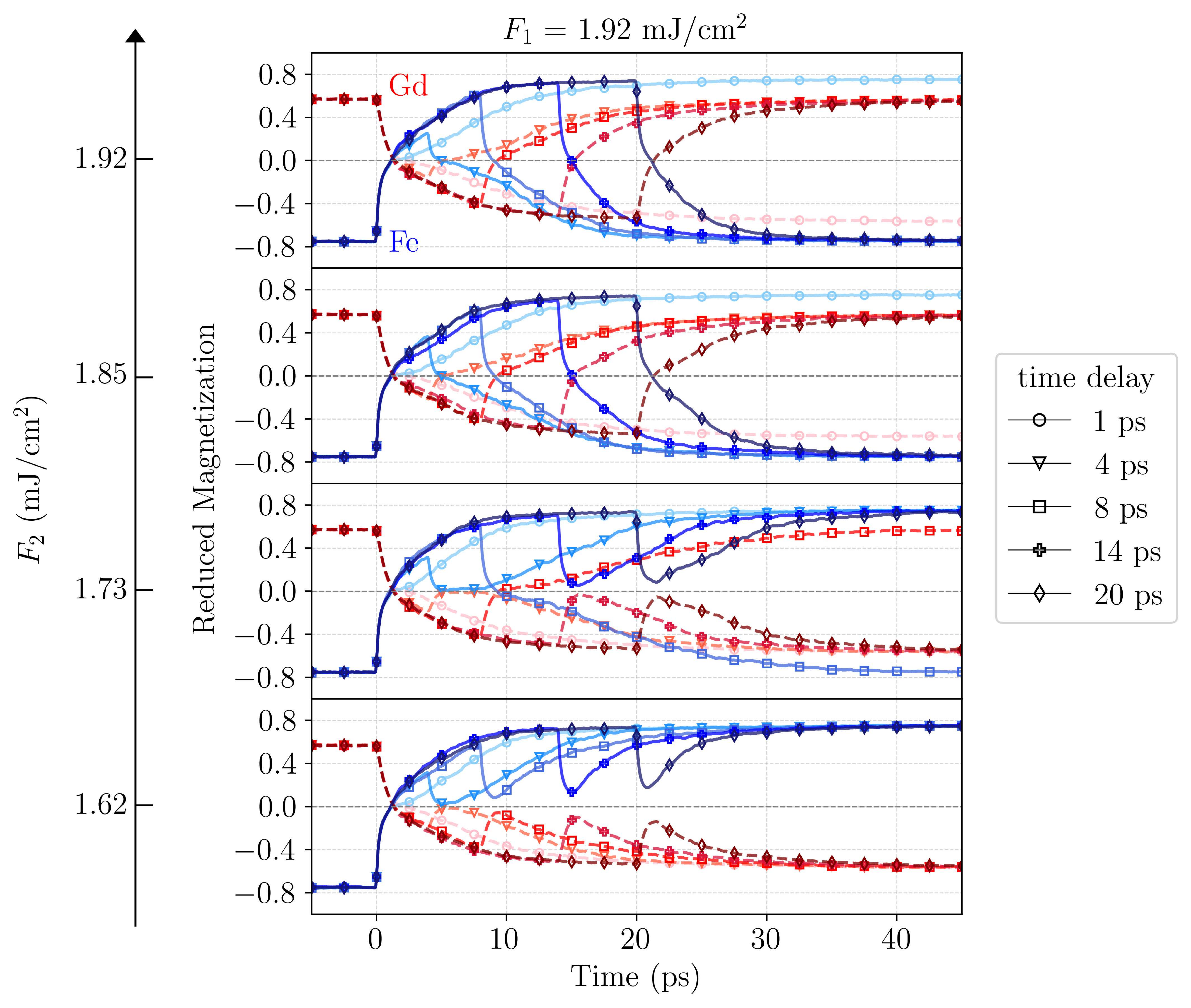}
    \caption{The magnetization dynamics of Fe and Gd sublattices for$F_{1} = \SI{1.92}{\milli\joule\per\centi\meter\textsuperscript{2}}$ just above the single-switching threshold.}
    \label{figure-s9}
\end{figure}

\begin{figure}[htb!]
    \centering
    \includegraphics[width=0.9\textwidth]{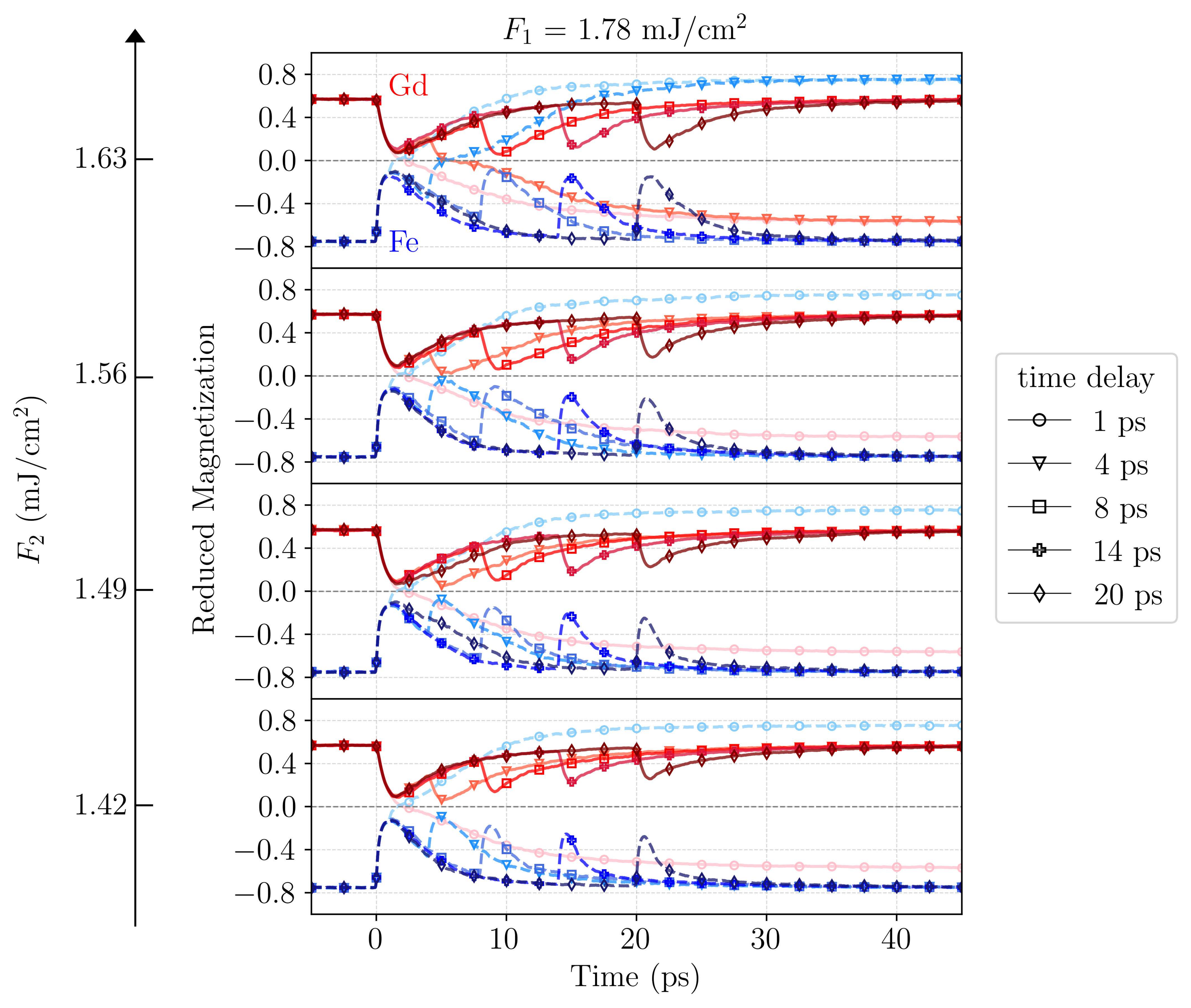}
    \caption{The magnetization dynamics of Fe and Gd sublattices for $F_{1} = \SI{1.78}{\milli\joule\per\centi\meter\textsuperscript{2}}$, below the single-switching threshold.}
    \label{figure-s10}
\end{figure}

Figure \ref{figure-s9} shows further examples of the temporal evolution of the Fe and Gd sublattice magnetizations for $F_{1} = \SI{1.92}{\milli\joule\per\centi\meter\textsuperscript{2}}$ from the map that is presented in Fig.\ 5 of the the main text. Figure \ref{figure-s10} presents the dynamics of each sub-lattice for $F_{1} = \SI{1.78}{\milli\joule\per\centi\meter\textsuperscript{2}}$.


\subsection*{Two-temperature model}
A two-temperature model was used in the simulations to calculate the temporal evolution
of the electron and lattice temperatures. The equations are presented in the main text.
The laser sources $P_{i}(t)$ follow the equation below:
\begin{equation}
    P_{i}(t) = D \cdot f (t)
\end{equation}

with $D = F_i \cdot dA$, where $F_i$ is the incident fluence and $dA$ represents the differential absorption in the middle of the GdFe layer as determined in Fig.\ \ref{figure-s1} . The function $f (t)$ is a Gaussian temporal profile:
\begin{equation}
 f (t) =\sqrt{\frac{\beta}{\pi t_{p}^{2}}} \cdot \exp{\left (-\beta \left(\frac{t-t_0}{t_p}\right)^2\right)}  
\end{equation}
where $\beta = 4\ln{2}$. We considered a pulse width of $t_{p} = \SI{100}{\femto\second}$ for both laser pulses and $t_{0}$ represents the time where each laser pulse reaches its maximum power. 
 In our simulations, we suppose no $z$ dependence of the laser power. Therefore we have, for the GdFe layer:
\begin{equation}
 D(z) = F \cdot dA(z) =\textrm{constant (for each fluence)}
\end{equation}
where $dA = \SI{13}{\mu\meter^{-1}}$ corresponds to the value of the differential absorption in the middle of the GdFe layer, provided by the model with vertical cooling \cite{shokr_steering_2019}.

\subsection*{ Different damping parameters}

\begin{figure}[htb!]
    \centering
    \includegraphics[width=0.6\textwidth]{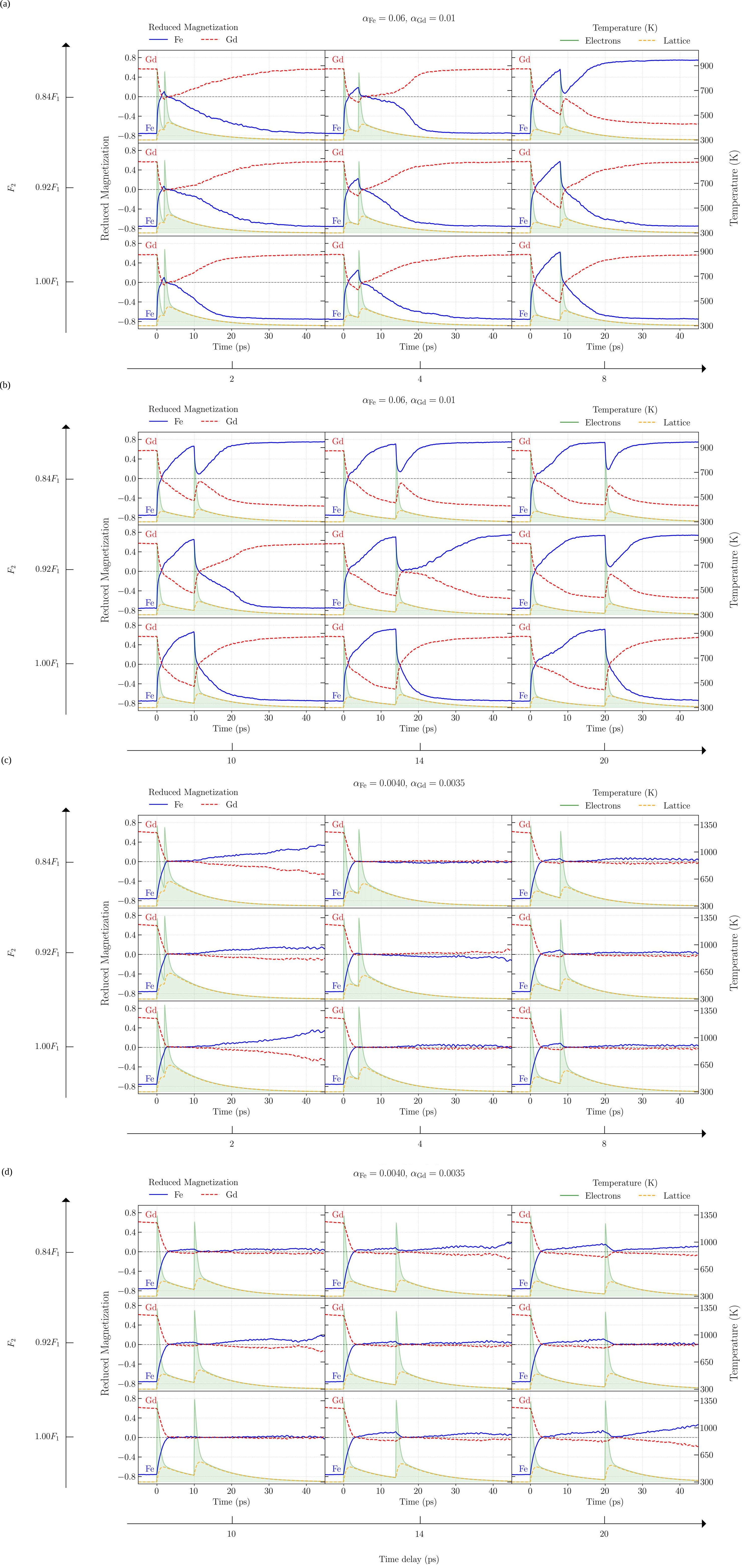}
    \caption{Simulated magnetization dynamics of Gd\textsubscript{26}Fe\textsubscript{74} alloy after a double-pulse excitation for different damping parameters where (a) $ \alpha_{Fe}$ = \SI{0.06}{} and $\alpha_{Gd}$ = \SI{0.01}{} for short time and (b) longer time delay, 
  (c) $\alpha_{Fe}$ = \SI{0.0040}{} and $\alpha_{Gd} $= \SI{0.035}{} for short time and (d) longer time delay.}
    \label{damping}
\end{figure}

Fig.\ \ref{damping} shows the comparison of the magnetization dynamics in the Gd\textsubscript{26}Fe\textsubscript{74} alloy after a double-pulse excitation calculated for different damping parameters, as detailed in the figure. Two sets of damping values are compared: $\alpha_{Fe}$ = \SI{0.06}{} and $\alpha_{Gd}$ = \SI{0.01}{} from Ref. \cite{jakobs_unifying_2021}
and $\alpha_{Fe}$ = \SI{0.0040}{} and $\alpha_{Gd}$ = \SI{0.035}{}, as recently reported in \cite{steinbach_exploring_2024}.
In all scenarios, the fluence of the first pulse matches the threshold  for single-pulse switching, with the second-pulse fluence set at a fixed proportion to that of the first pulse ($F_2$/$F_1$ =  \SI{0.84}{}, \SI{0.92}{}, \SI{1.00}{}).
Simulations show that the double-switching effect seen with higher damping disappears entirely at the lower values of damping. Intermediate damping values confirm a smooth transition between these two cases.

\bibliographystyle{apsrev4-2}
\bibliography{Ref}